\documentclass[11pt,graphicx,math,amsmath,amssymb]{article}
\usepackage{hyperref}

\def\be{\begin{equation}}
\def\ee{\end{equation}}
\def\bea{\begin{eqnarray}}
\def\eea{\end{eqnarray}}

\def\IR{\relax{\rm I\kern-.18em R}}
\def\binomial#1#2{\left(\,{\buildrel
{\raise4pt\hbox{$\displaystyle{#1}$}}\over
{\raise-6pt\hbox{$\displaystyle{#2}$}}}\,\right)}

\def\[{\lfloor{\hskip 0.35pt}\!\!\!\lceil}
\def\]{\rfloor{\hskip 0.35pt}\!\!\!\rceil}

\def\a{\alpha}
\def\b{\beta}
\def\d{\delta}
\def\e{\epsilon}
\def\f{\phi}
\def\g{\gamma}
\def\h{\eta}

\def\k{\kappa}
\def\l{\lambda}
\def\m{\mu}
\def\n{\nu}
\def\o{\omega}

\def\q{\theta}
\def\r{\rho}
\def\s{\sigma}
\def\t{\tau}

\def\F{\Phi}
\def\G{\Gamma}

\def\S{\Sigma}

\def\cd{{\cal D}}

\def\cl{{\cal L}}
\def\cm{{\cal M}}
\def\cn{{\cal N}}

\def\bo{{\raise.15ex\hbox{\large$\Box$}}}               
\def\TH{{\raise.2ex\hbox{$\displaystyle \bigodot$}\mskip-4.7mu \llap H
\;}}
\def\face{{\raise.2ex\hbox{$\displaystyle \bigodot$}\mskip-2.2mu \llap
{$\ddot
        \smile$}}}                                      

\def\Hat#1{\widehat{#1}}                        
\def\Bar#1{\overline{#1}}                       
\def\leftrightarrowfill{$\mathsurround=0pt \mathord\leftarrow \mkern-6mu
        \cleaders\hbox{$\mkern-2mu \mathord- \mkern-2mu$}\hfill
        \mkern-6mu \mathord\rightarrow$}
\def\dvec#1{\vbox{\ialign{##\crcr
        \leftrightarrowfill\crcr\noalign{\kern-1pt\nointerlineskip}
        $\hfil\displaystyle{#1}\hfil$\crcr}}}           

\catcode`@=11
\def\un#1{\relax\ifmmode\@@underline#1\else
        $\@@underline{\hbox{#1}}$\relax\fi}
\catcode`@=12

\def\fracm#1#2{\hbox{\large{${\frac{{#1}}{{#2}}}$}}}
\def\frac#1#2{{\textstyle{#1\over\vphantom2\smash{
        \hbox{$\scriptstyle{#2}$}}}}}                   
\def\sfrac#1#2{{\vphantom1\smash{\lower.5ex\hbox{\small$#1$}}\over
        \vphantom1\smash{\raise.4ex\hbox{\small$#2$}}}} 
\def\bfrac#1#2{{\vphantom1\smash{\lower.5ex\hbox{$#1$}}\over
        \vphantom1\smash{\raise.3ex\hbox{$#2$}}}}       
\def\afrac#1#2{{\vphantom1\smash{\lower.5ex\hbox{$#1$}}\over#2}}    

\newskip\humongous \humongous=0pt plus 1000pt minus 1000pt
\def\caja{\mathsurround=0pt}
\def\eqalign#1{\,\vcenter{\openup2\jot \caja
        \ialign{\strut \hfil$\displaystyle{##}$&$
        \displaystyle{{}##}$\hfil\crcr#1\crcr}}\,}
\newif\ifdtup

\def\pp{{\mathchoice
          {
              \kern 1pt%
              \raise 1pt
              \vbox{\hrule width5pt height0.4pt depth0pt
                    \kern -2pt
                    \hbox{\kern 2.3pt
                          \vrule width0.4pt height6pt depth0pt
                          }
                    \kern -2pt
                    \hrule width5pt height0.4pt depth0pt}%
                    \kern 1pt
           }
            {
              \kern 1pt%
              \raise 1pt
              \vbox{\hrule width4.3pt height0.4pt depth0pt
                    \kern -1.8pt
                    \hbox{\kern 1.95pt
                          \vrule width0.4pt height5.4pt depth0pt
                          }
                    \kern -1.8pt
                    \hrule width4.3pt height0.4pt depth0pt}%
                    \kern 1pt
            }
            {
              \kern 0.5pt%
              \raise 1pt
              \vbox{\hrule width4.0pt height0.3pt depth0pt
                    \kern -1.9pt  
                    \hbox{\kern 1.85pt
                          \vrule width0.3pt height5.7pt depth0pt
                          }
                    \kern -1.9pt
                    \hrule width4.0pt height0.3pt depth0pt}%
                    \kern 0.5pt
            }
            {
              \kern 0.5pt%
              \raise 1pt
              \vbox{\hrule width3.6pt height0.3pt depth0pt
                    \kern -1.5pt
                    \hbox{\kern 1.65pt
                          \vrule width0.3pt height4.5pt depth0pt
                          }
                    \kern -1.5pt
                    \hrule width3.6pt height0.3pt depth0pt}%
                    \kern 0.5pt
            }
        }}

  \def\mm{{\mathchoice
                       {
                             \kern 1pt
               \raise 1pt    \vbox{\hrule width5pt height0.4pt depth0pt
                                  \kern 2pt
                                  \hrule width5pt height0.4pt depth0pt}
                             \kern 1pt}
                       {
                            \kern 1pt
               \raise 1pt \vbox{\hrule width4.3pt height0.4pt depth0pt
                                  \kern 1.8pt
                                  \hrule width4.3pt height0.4pt depth0pt}
                             \kern 1pt}
                       {
                            \kern 0.5pt
               \raise 1pt
                            \vbox{\hrule width4.0pt height0.3pt depth0pt
                                  \kern 1.9pt
                                  \hrule width4.0pt height0.3pt depth0pt}
                            \kern 1pt}
                       {
                           \kern 0.5pt
             \raise 1pt  \vbox{\hrule width3.6pt height0.3pt depth0pt
                                  \kern 1.5pt
                                  \hrule width3.6pt height0.3pt depth0pt}
                           \kern 0.5pt}
                       }}

\def\dslash{\not{\hbox{\kern-2pt $\partial$}}}
\def\Dslash{\not{\hbox{\kern-4pt $D$}}}
\def\pslash{\not{\hbox{\kern-2.3pt $p$}}}
 \newtoks\slashfraction
 \slashfraction={.13}
 \def\slash#1{\setbox0\hbox{$ #1 $}
 \setbox0\hbox to \the\slashfraction\wd0{\hss \box0}/\box0 }

\font\ro=cmsy10                          
\def\kcr{{\hbox{\ro \char'170}}}                
\def\ktl{{\hbox{\ro \char'170}}}        
\def\ktr{{\hbox{\ro \char'170}}}        
\def\kbl{{\hbox{\ro \char'170}}}        
\def\kbr{{\hbox{\ro \char'170}}}        

\def\plpl{\raise-2pt\hbox{$\raise3pt\hbox{$_+$}\hskip-6.67pt\raise0.0pt
\hbox{$^+$}\hskip 0.01pt$}}
\def\mimi{\raise-2pt\hbox{$\raise3pt\hbox{$_-$}\hskip-6.67pt\raise0.0pt
\hbox{$^-$}\hskip 0.01pt$}}

\parskip=\medskipamount                 
\thicklines                         

\def\border{                                            
        \setlength{\unitlength}{1mm}
        \newcount\xco
        \newcount\yco
        \xco=-21
        \yco=12
        \begin{picture}(140,0)
        \put(\xco,\yco){$\ktl$}
        \advance\yco by-1
        {\loop
        \put(\xco,\yco){$\kcr$}
        \advance\yco by-2
        \ifnum\yco>-240
        \repeat
        \put(\xco,\yco){$\kbl$}}
        \xco=158
        \yco=12
        \put(\xco,\yco){$\ktr$}
        \advance\yco by-1
        {\loop
        \put(\xco,\yco){$\kcr$}
        \advance\yco by-2
        \ifnum\yco>-240
        \repeat
        \put(\xco,\yco){$\kbr$}}
        \put(-20,13){\tiny **University of Maryland * Center for String and
         Particle  Theory* Physics Department***University of Maryland *Center
        for String and Particle  Theory** }
        \put(-20,-241.5){\tiny **University of Maryland * Center for String and
         Particle  Theory* Physics Department***University of Maryland *Center
        for String and Particle  Theory** }
        \end{picture}
        \par\vskip-8mm}
\def\headpic{                                           
        \indent
        \setlength{\unitlength}{.4mm}
        \thinlines
        \par
        \begin{picture}(29,16)
        \put(165,16){\line(1,0){4}}
        \put(170,16){\line(1,0){4}}
        \put(180,16){\line(1,0){4}}
        \put(175,0){\line(1,0){4}}
        \put(180,0){\line(1,0){4}}
        \put(185,0){\line(1,0){4}}
        \put(169,0){\line(0,1){16}}
        \put(170,0){\line(0,1){16}}
        \put(179,0){\line(0,1){16}}
        \put(180,0){\line(0,1){16}}
        \put(184,0){\line(0,1){16}}
        \put(185,0){\line(0,1){16}}
        \put(169,16){\oval(8,32)[bl]}
        \put(170,16){\oval(8,32)[br]}
        \put(179,0){\oval(8,32)[tl]}
        \put(185,0){\oval(8,32)[tr]}
        \end{picture}
        \par\vskip-6.5mm
        \thicklines}
\def\title#1#2#3#4{\border\headpic {\hbox to\hsize{#4 \hfill UMDEPP #3}}\par
        \begin{center} \vglue .5in {\large\bf #1}\\[.6in]
        {#2}\\[.1in] {\it Department of Physics and Astronomy}\\
        {\it University of Maryland, College Park, MD 20742}\\[1.5in]
        {\bf ABSTRACT}\\[.1in] \end{center} \begin{quotation}}  
\def\Title#1#2#3#4#5#6#7{\border\headpic
        {\hbox to\hsize{#7 \hfill UMDEPP #6}}\par
        \begin{center} \vglue .4in {\large\bf #1}\\[.4in]
        {#2}\\[.1in] {\it Department of Physics and Astronomy}\\
        {\it University of Maryland, College Park, MD 20742}\\[.1in]
        {#3}\\[.1in] {\it {#4}}\\ {\it {#5}}\\[.5in] {\bf ABSTRACT}\\[.1in]
        \end{center} \begin{quotation}}                 
\def\endtitle{\end{quotation}\newpage}                  

\def\qd{{\kern0.5pt
                   q \kern-5.05pt \raise5.8pt\hbox{$\textstyle.$}\kern
0.5pt}}

\begin{document}

\def\gfrac#1#2{\frac {\scriptstyle{#1}}
        {\mbox{\raisebox{-.6ex}{$\scriptstyle{#2}$}}}}
\def\gg{{\hbox{\sc g}}}
\border\headpic {\hbox to\hsize{December 1, 2003 \hfill
{UMD-PP 03-071}}}
{\hbox to\hsize{~\hfill hep-th/0312040}}
\par

\setlength{\oddsidemargin}{0.3in}
\setlength{\evensidemargin}{-0.3in}

\begin{center}
\vglue .10in

{\large\bf{$\cal{M}$-theory on $Spin(7)$ Manifolds, Fluxes\\ and
3D, ${\cal N}$=1 Supergravity}}
\\[1cm]

Melanie Becker\footnote{\href{mailto:melanieb@physics.umd.edu}{melanieb@physics.umd.edu},
${}^2$\href{mailto:dragos@physics.umd.edu}{dragos@physics.umd.edu},
 ${}^3$\href{mailto:gatess@wam.umd.edu}{gatess@wam.umd.edu},\\
  ${}^4$\href{mailto:ldw@physics.umd.edu}{ldw@physics.umd.edu},
  ${}^5$\href{mailto:williem@physics.umd.edu}{williem@physics.umd.edu},
  ${}^6$\href{mailto:ferrigno@physics.umd.edu}{ferrigno@physics.umd.edu}},
Drago\c{s} Constantin${}^2$, \\
S. James Gates, Jr.${}^3$,
William D. Linch, III${}^4$,
Willie Merrell${}^5$,\\
and J. Phillips${}^6$
\\[1cm]
{\it Center for String and Particle Theory\\
Department of Physics, University of Maryland\\
College Park, MD 20742-4111 USA}
\\[2cm]

{\bf ABSTRACT}\\[.01in]

\end{center}
We calculate the most general causal ${\cal N}=1$
three-dimensional, gauge invariant action coupled to matter in
superspace and derive its component form using Ectoplasmic
integration theory. One example of such an action can be obtained
by compactifying ${\cal M}$-theory on a $Spin(7)$ holonomy
manifold taking non-vanishing fluxes into account. We show that
the resulting three-dimensional theory is in agreement with the
more general construction. The scalar potential resulting from
Kaluza-Klein compactification stabilizes all the moduli fields
describing deformations of the metric except for the radial
modulus. This potential can be written in terms of the
superpotential previously discussed in the literature.


\def\c{\chi}

\newpage


\setcounter{page}{2}


\renewcommand{\thesection}{\arabic{section}}


\section{Introduction}
\label{intro}

\setcounter{equation}{0}
\renewcommand{\theequation}{\thesection.\arabic{equation}}


$~~~$Component formulations of supergravity in various dimensions
with extended supersymmetry have been known for a long time.  In
general, the extended supergravities can be obtained by
dimensional reduction and truncation of higher dimensional
supergravities.  For example, a four-dimensional supergravity
with ${\cal N} = 1 $ supersymmetry leads to a three-dimensional
supergravity with ${\cal N}=2$ supersymmetry after
compactification.  For this reason the component form of
three-dimensional ${\cal N}=2$ supergravity is known.  Although
there has been much activity in three dimensions
\cite{Rocek:1986bk, vanNieuwenhuizen:1985cx, Karlhede:1987qd,
Karlhede:1987qf, Park:1999cw, Ivanov:2000tz, Zupnik:1999tf,
Zupnik:1997nn, Alexandre:2003qb}, there is no general {\it off-shell} component or superspace
formulation of three-dimensional ${\cal N}=1$ supergravity in the
literature. There are, however, on-shell realisations with ${\cal N} \geq 1$ given
 in \cite{deWit:1993up, deWit:2003ja}.\footnote{We thank Michael Haack for bringing these references to our attention.} The ${\cal N}=1$ theory cannot be obtained by dimensional
reduction from a four-dimensional theory and requires a formal
analysis. One of the goals of this paper is to derive the most
general off-shell three-dimensional ${\cal N}=1$ supergravity action
coupled to an arbitrary number of scalars and $U(1)$ gauge fields.

Although the off-shell formulation of ${\cal N}=1$
three-dimensional supergravity has been around since 1979
\cite{Brown:1979ma}, there has been little work done on
understanding this theory with the same precision and detail of
the minimal supergravity in four dimensions.  The spectrum of the
${\cal N}=1$ three-dimensional supergravity theory consists of a
dreibein, a Majorana gravitino and a single real auxiliary scalar
field. Since our formal analysis yields an off-shell formulation,
we can freely add distinct super invariants to the action. The
resulting theory corresponds to a non-linear sigma model and
copies of $U(1)$ gauge theories coupled to supergravity.  We will
present the complete superspace formulation in the hope that the
presentation will familiarize the reader with the techniques
required to reach our goals.

Three-dimensional supergravity theories can be obtained from
compactifications of ${\cal M}$-theory with non-vanishing fluxes.\footnote{Some
early references about warped compactifications of string theory are \cite{Strominger:1986uh, deWit:1987xg}}
Such theories were first considered in \cite{Becker:1996gj,
Sethi:1996es}, and later on generalized to the Type IIB theory in
\cite{Dasgupta:1999ss}.

In order to compactify while preserving
supersymmetry, we must consider internal manifolds that admit
covariantly constant spinors. Once the background metric is
chosen, the shape and size of the internal manifold can still be
deformed, which leads to scalar fields in the low-energy
effective supergravity theory, the so called moduli fields. If
the compactified theory contains no scalar potential, the moduli
fields can take any possible values and the theory loses
predictive power. However, it was realized in
\cite{Dasgupta:1999ss, Taylor:1999ii, Gukov:1999ya, Haack:2001jz,
Giddings:2001yu, Berg:2002es} that for string theory and ${\cal M}$-theory
compactifications with non-vanishing fluxes a scalar potential
emerges, which stabilizes many of the moduli fields. Therefore,
predictions for the coupling constants can be made. In order to
connect the compactified ${\cal M}$-theory with the superspace
formulation it is necessary to integrate out the auxiliary fields
in the latter theory by using the algebraic equations of motion.
This process shows how the scalar potential is naturally written
in terms of the superpotential.

For various reasons three-dimensional compactifications of $\cal
M$-theory with minimal supersymmetry are specially interesting.
First, it has been suggested that compactifications to three
dimensions with ${\cal N}=1$ supersymmetry could naturally
explain a small cosmological constant in four-dimensions
\cite{Witten:1994cg, Witten:1995rz}.

Second, recall that it was shown in \cite{Grana:2000jj,
Gubser:2000vg} that some particular type of compactifications of
${\cal M}$-theory to three dimensions with ${\cal N}=2$
supersymmetry are the supergravity dual of the four-dimensional
confining gauge theory found in \cite{Klebanov:2000hb}. A duality
of this type is of practical interest since calculations in the
strongly coupled gauge theory can be performed in the dual weakly
coupled supergravity theory in the spirit of the AdS/CFT
correspondence found in \cite{Maldacena:1997re}. From the point
of view of supersymmetry, 3D, ${\cal N}=1$ gauge theories are
similar to 4D, ${\cal N}=0$ theories. So insight into 4D, ${\cal
N}=0$ gauge theories could be gained from studying the
supergravity duals of 3D, ${\cal N}=1$ theories. This is a
developing area and a complete discussion is beyond the scope of
this paper.  Instead, we refer the interested reader to
\cite{Becker:1996gj, Sethi:1996es, Dasgupta:1999ss, Grana:2000jj,
Gubser:2000vg, Klebanov:1998hh, Klebanov:2000nc, Klebanov:2000hb,
Polchinski:2000uf}.

In order to compactify ${\cal M}$-theory to three dimensions,
keeping only the minimal supersymmetry, we require an
eight-dimensional internal manifold admitting a single
covariantly constant spinor. In the case of a compact Riemannian
internal space, this leads us uniquely to manifolds with
$Spin(7)$ holonomy. Early papers about compactification of ${\cal
M}$-theory on manifolds with exceptional holonomy are e.g.
\cite{Shatashvili:1994zw, Papadopoulos:1995da}. The constraints
on the fluxes following from supersymmetry for compactifications
of $\cal M$-theory on $Spin(7)$ holonomy manifolds with
background fields were first derived in \cite{Becker:2000jc}. It
was later shown in \cite{Acharya:2002vs} and \cite{Becker:2002jj},
that these constraints can be derived from a superpotential,
whose explicit form is in accordance with the conjecture made in
\cite{Gukov:1999gr}. Several interesting examples and aspects of
these compactifications have been discussed in the literature
(see e.g. \cite{Gukov:2001hf, Cvetic:2001pg} and references
therein). In this paper, we shall calculate the Kaluza-Klein
compactification of ${\cal M}$-theory on a $Spin(7)$ holonomy
manifold with non-vanishing fluxes.  Our calculation is similar
to that of \cite{Haack:2001jz} which was done in the context of
${\cal M}$-theory compactifications on conformally Calabi-Yau
four-folds. We will see that the resulting scalar potential leads
to the stabilization of all the moduli fields corresponding to
deformations of the internal manifold except the radial modulus.
This scalar potential can be expressed in terms of the
superpotential which has appeared previously in the literature
\cite{Acharya:2002vs, Becker:2002jj, Gukov:1999gr}.

This paper is organized as follows: Section \ref{3d-sugra}, is
devoted to the introduction of the geometry and dynamics of
three-dimensional ${\cal N}=1$ supergravity coupled to matter. In
section \ref{supergeometry}, we present the algebra of
supercovariant derivatives which describes the superspace
geometry. We then discuss Ectoplasmic integration, the technique
used to calculate the density projector, which is required to
integrate over curved supermanifolds. In section
\ref{super-3forms}, we solve the Bianchi identities for a super
three-form subject to the given algebra required for Ectoplasmic
integration. In section \ref{ectoplasmic}, we detail the use of
Ectoplasm to calculate the density projector. In section
\ref{component-formalism}, we complete the supergravity analysis
by first deriving the component fields and then calculating the
component action. We end the analysis by giving the supersymmetry
transformations for the component fields and putting the
component action on shell i.e. we remove the auxiliary fields by
their algebraic equations of motion. Section
\ref{compactification} is devoted to the compactification of
${\cal M}$-theory on a $Spin(7)$ holonomy manifold. We start in
section \ref{zero-flux} by compacifying without background
fluxes. In section \ref{non-zero-flux}, we take the background
fluxes into account and derive the complete form of the bosonic
part of the action. We show that this action is a special case of
our findings in the general construction of section
\ref{3d-sugra}. In section \ref{open-questions}, we give a
summary of our results and comment on the physics implied by the
explicit form of the scalar potential. We conclude this section
with some open questions and directions for future investigations
suggested by our findings.  Finally, details related to our
calculations are contained in the appendices.  Appendix
\ref{conventions} contains our notations and conventions.
Appendix \ref{fierz} describes the derivation of the
three-dimensional Fierz identities. In appendix \ref{3D-algebra},
we derive the closure of the three-dimensional super covariant
derivative algebra.  In appendix \ref{appendix-spin7}, we provide
relevant aspects related to manifolds with $Spin(7)$ holonomy. In
appendix \ref{appendix-dimensional}, we perform the Kaluza-Klein
compactification of the eleven-dimensional Einstein-Hilbert term
on a $Spin(7)$ holonomy manifold.


\section{Minimal $3D$ Supergravity Coupled to Matter}
\label{3d-sugra}

\setcounter{equation}{0}
\renewcommand{\theequation}{\thesection.\arabic{equation}}


$~~~$Using the Ectoplasmic Integration theorem we derive the
component action for the general form of supergravity coupled to
matter.  The matter sector includes $U(1)$ gauge fields and a
non-linear sigma model.


\subsection{Supergeometry}
\label{supergeometry}

\setcounter{equation}{0}
\renewcommand{\theequation}{\thesubsection.\arabic{equation}}


$~~~$Calculating component actions from manifestly supersymmetric
supergravity descriptions is a complicated process.  Knowing the
supergravity density projector simplifies this process.  The
density projector arises from the following observation.  Every
supergravity theory that is known to possess an off-shell
formulation can be be shown to obey an equation of the form:
\bea
\int d^{D} x \, d^{\cal N}  \theta \, \, {\rm E}{}^{-1}\cl
~=~ \int d^{D}x ~ {\rm e}{}^{-1}(\cd^{\cal N} \cl|) \,,
\eea
for a superspace with space-time dimension $D$, and fermionic dimension
$\cal N$.  E${}^{-1}$ is the super determinant of the super
frame fields E${}_{A}{}^{M}$, $\cd^{\cal N}$ is a differential
operator called the supergravity density projector, and the symbol $|$ denotes taking the anti-commuting coordinate to zero.  This relation has been dubbed the Ectoplasmic Integration Theorem and
shows us that knowing the form of the density projector allows us to evaluate the component structure of any Lagrangian just by evaluating $(\cd^{\cal N} \cl|)$.  Thus, the problem of finding components for supergravity is relegated to computing the density projector.

Two well defined methods for calculating the density projector exist in the literature.  The first method is based on super $p$-forms and the Ethereal
Conjecture.  This conjecture states that in all supergravity theories, the topology of the superspace is determined {\it {solely}} by its bosonic
submanifold.  The second method is called the ectoplasmic normal coordinate expansion \cite{Grisaru:1997ub, Gates:1997ag}, and explicitly calculates the density projector.  The normal coordinate expansion provides a proof of the ectoplasmic integration theorem.  
Both of these techniques rely heavily on the algebra of
superspace supergravity covariant derivatives.  The covariant derivative algebra for three dimensional supergravity was first given in \cite{Brown:1979ma}.  In this paper, we have modified the original algebra by coupling it to $n ~ U(1)$, gauge fields:\footnote{We do not consider non-abelian gauged supergravity because the compactifications of ${\cal M}$-theory on $Spin(7)$ manifolds that we consider lead to abelian gauged supergravities.}
\bea
\label{algebra}
[ \nabla_{\a} ~,~ \nabla_{\b} \}  &=&
(\g^c)_{\a \b} ~ \nabla_c ~-~ (\g^c)_{\a \b}R \, {\cal
M}_c \,, \cr
~[ \nabla_\a ~,~ \nabla_b \}  &=& ~ +\frac 12(\g_b)_{\a}
{}^{\d} R \nabla_{\d} ~+~[-2(\g_b)_{\a}{}^{\d}  \S_{\d}
{}^d ~-~  \fracm 23 (\g_b\g^d)_{\a}{}^{\e} ( \nabla_{\e}
R )] {\cal M}_d ~\cr &~&~+~ (\nabla_{\a} R ) {\cal M}_b
+ \frac 13(\g_b)_\a^{~\b}W_\b^I t_I \,,
\cr
[ \nabla_a ~,~ \nabla_b \}  &=&   +2 \e_{abc} [~- \S^{
\a c} - \fracm 13 (\g^c)^{\a \b} (\nabla_{\b} R) ~]
\nabla_{\a}  \cr
&~&+  \e_{abc}[~  {\Hat {\cal R}}{}^{cd} ~+~ \fracm 23
\eta^{cd} (-2\nabla^2 R ~-~ \fracm 32 R^2 )~] {\cal
M}_d\cr &~&+\frac 13\e_{abc}(\g^{c})_\b^{~\d}\nabla^\b
W_\d^I t_I  \,, \eea
where
\bea
\label{secondfs}
{\Hat {\cal R}}{}^{ab} - {\Hat {\cal R}}{}^{ba} =
\eta_{ab}{\Hat {\cal R}}{}^{ab} = (\g_d)^{\a \b} \S_{
\b}{}^d = 0 \,,
~~~~~~~~~~\nonumber\\
\nabla_\a \S_\b^{~f} = -\frac 14 (\g^e)_{\a\b}\Hat
{\cal R}_e^{~f} +\frac 16[C_{\a\b}\h^{fd}+\frac 12
\e^{fde}(\g_e)_{\a\b}]\nabla_d R\,,\cr
\nabla^\d W_\d^I = 0 \,.~~~~~~~~~~
~~~~~~~~~~~~~~~~~~~~
\eea
The superfields $R, ~\S_{\a}{}^b$ and $\Hat{\cal R}^{ab}$
are the supergravity field strengths, and  $W_\a^I$ are
the $U(1)$ super Yang-Mills fields strengths. $t_I$ are the
$U(1)$ generators with $I=1\dots n$.  $\cm_a$ is the 3D Lorentz generator.  Our convention for the action of $\cm_a$ is given in appendix \ref{3d-sugra-conventions}.  An explicit verification of the algebra (\ref{algebra}) is performed in appendix \ref{3D-algebra}, where it is shown that the algebra closes off-shell.

In this paper, we choose to use the method of ectoplasmic integration.  The following three sections outline the implementation of this procedure.


\subsection{Closed Irreducible 3D, $\cal N$ = 1 Super 3-forms}
\label{super-3forms}

\setcounter{equation}{0}
\renewcommand{\theequation}{\thesubsection.\arabic{equation}}


${~~~}$Indices of topological significance in a
D-dimensional space-time manifold can be calculated from the
integral of closed but not exact D-forms.  The Ethereal Conjecture
suggests that this reasoning should hold for superspace.  Thus, in order to use the Ethereal Conjecture \cite{Gates:1998hy}, we must first have the description of a super 3-form field strength.  In this section, we derive the super 3-form associated with the covariant derivative algebra (\ref{algebra}).

We start with the general formulas for the super 2-form potential and super 3-form field strength.  A super 2-form $\G_2$ has the
following gauge transformations:
\bea
\label{FBianchi1}
\d\G_{AB} = \nabla_{[A}K_{B)}-\frac 12 T_{[AB)}^{
~~~E} K_{E}~~,
\eea
which expresses the fact that the gauge variation of
the super 2-form is the super-exterior derivative of
a super 1-form $K_1$.   The field strength $G_3$ is
the super exterior derivative of $\G_2$:
\bea
G_{ABC} = \frac 12\nabla_{[A}\G_{BC)}-\frac 12 T_{[AB|
}^{~~~~E}\G_{E|C)}~~.
\eea
We have a few comments about the notation in these expressions.  First, upper case roman indices are super vector
indices which take values over both the spinor and vector indices.  Also, letters from the beginning(middle) of the alphabet refer to flat(curved) indices.  Finally, the symmetrization symbol $[)$ is a graded symmetrization.  A point worth noting here is the that the superspace torsion appears explicitly
in these equations.  This means that the super-form is intimately
related to the type of supergravity that we are using.
The appearance of the torsion in these expressions is
{\it {not}} peculiar to supersymmetry.  Whenever forms
are referred to using a non-holonomic basis this phenomenon
occurs.

A super-form is a highly reducible representation
of supersymmetry.  Therefore, we must impose certain constraints
on the field strength to make it an irreducible
representation of supersymmetry.  In general, there are many types of constraints that we can set.  Different constraints have specific consequences.  A conventional constraint implies that one piece of the potential is related to another.  In this case if we set the conventional constraint:
\bea
G_{\a\b\g} = \frac 12 \nabla_{(\a}\G_{\b\g)}-\frac 12(\g^c
)_{(\a\b|}\G_{c|\g)}=0~~,
\eea
we see that the potential $\G_{c\a}$ is now related to the spinorial
derivative of the potential $\G_{\a\b}$.  This constraint eliminates six superfield degrees of freedom.

Since $G_3$ is the exterior derivative of a super 2-form it
must be closed, i.e. its exterior derivative $F_4$ must vanish.
This constitutes a set of Bianchi identities:
\bea
\label{FBianchi}
F_{ABCD}=\frac 1{3!}\nabla_{[A}G_{BCD)}-\frac 14T_{[AB|}^{~~~~
E}G_{E|CD)}=0~~.
\eea
Once a constraint has been set, these Bianchi identities are no longer identities.  In fact, the consistency of the Bianchi identities after a constraint has been imposed implies an entire set of constraints.  By solving the Bianchi identities with respect to the conventional constraint, we can completely  determine the irreducible super 3-form field strength.  Since we have set $G_{\a\b\g}=0$, it is easiest
to solve $F_{\a\b\g\d}=0$ first:
\bea
F_{\a\b\g\d} &=& \frac 16 \nabla_{(\a}G_{\b\g\d)}-\frac 14
T_{(\a\b|}^{~~~E}G_{E|\g
\d)}
=-\frac 14(\g^e)_{(\a\b|}G_{e|\g\d)}~~.
\eea
To solve this equation, we must write out the Lorentz irreducible parts of $G_{a\b\g}$.  We first convert the last two spinor indices on $G_{e\g\d}$ to a vector index by contracting with the gamma matrix: $G_{e\g\d}=(\g^f)_{\g\d}G_{ef}$.  Further, $G_{a\b\g}=G_{\b\g a}$ implies that $G_{ab}$ is a symmetric tensor, so we make the following decomposition: $G_{ab} = \Bar G_{ab} +\frac 13\h_{ab}G^d_{~d}$, where the bar on $\Bar G_{ab}$ denotes tracelessness.  With this decomposition, the Bianchi identity now reads:
\bea
F_{\a\b\g\d}=-\frac 14(\g^e)_{(\a\b}(\g^f)_{\g\d)}\Bar G_{ef}=0~~,
\eea
where the term containing $G^d{}_d$ vanishes exactly.  The symmetric traceless part of this gamma matrix structure does not vanish, so we are forced to set $\Bar G_{ab}=0$.  Thus, our conventional constraint implies the further constraint $G_{a\b\g}=(\g_a)_{\b\g}G$.  The next Bianchi identity reads:
\bea
F_{\a\b\g d} = \frac 12 \nabla_{(\a}G_{\b\g)d}
-\frac 1{3!}\nabla_dG_{(\a\b\g)}
-\frac 12 T_{(\a\b|}^{~~~E}G_{E|\g)d}
+\frac 12 T_{d(\a|}^{~~~E}G_{E|\b\g)}~~.
\eea
Using our newest constraint and substituting the torsions we
have:
\bea
\label{Fbianchi2}
F_{\a\b\g d} &=& \frac 12 (\g_d)_{(\b\g}\nabla_{\a)}G
+\frac 12 (\g^e)_{(\a\b|}G_{\g)ed}\cr
&=& \frac 12 (\g_d)_{(\b\g}\nabla_{\a)}G
+\frac 12 (\g^e)_{(\a\b|}\e_{ed}^{~~a}[(\g_a)_{\g)}^{~~\d}
G_{\d}+\hat G_{\g)a}]~~,
\eea
here we have replaced the antisymmetric vector indices with a Levi-Cevita tensor via; $G_{\g ed}=\e_{ed}^{~~a}G_{\g a}$, and further decomposed $G_{\g a}$ into spinor and gamma traceless parts; $G_{\g a}=(\g_a)_\g{}^\b G_\b+\hat G_{\g a}$, respectively.  Contracting (\ref{Fbianchi2}) with $\e_{cbe}
(\g^e)^{\a\b}\d_\s^{~\g}$ implies $G_\s = \nabla_\s
G$.  Substituting this result back into (\ref{Fbianchi2}) implies
that $\hat G_\g{}^a=0$.  Thus, we have derived another constraint on
the field strength:
\bea
G_{\a bc} = \e_{bc}^{~~a}(\g_a)_\a^{~\s}\nabla_\s G~~.
\eea
The third bianchi identity will completely determine the super
3-form:
\bea
F_{\a\b cd} &=&  \nabla_{(\a}G_{\b)cd} +\nabla_{[c}G_{d]
\a\b} - T_{\a\b}^{~~E}G_{Ecd} -T_{cd}^{~~E}G_{E\a\b}
-T_{(\a|[c|}^{~~~~E}G_{E|d]\b)}\cr
&=&\e_{cd}^{~~e}(\g_e)_{(\a}^{~~\s}\nabla_{\b)}\nabla_\s G
+(\g_{[d})_{\a\b}\nabla_{c]}G
-(\g^e)_{\a\b}G_{ecd}
+(\g_{[c}\g_{d]})_{\a\b}RG~.
\eea
Note that $G_{\a b \g}= -(\g_b)_{\a\g}G$.  Contracting
with $(\g_b)^{\a\b}$ yields the following equation for
the vector 3-form:
\bea
G_{bcd}=2\e_{bcd}[\nabla^2 G + RG]~~.
\eea
The final two bianchi identities are consistency checks
and vanish identically.
\bea
F_{\a bcd} = \frac 1{3!}\nabla_\a G_{[bcd]}-\frac 12
\nabla_{[b}G_{cd]\a}-\frac 12 T_{\a [b|}^{~~~E}G_{E|cd]}
+\frac 12 T_{[bc|}^{~~~E}G_{E|d]\a}=0\cr
F_{abcd}=\frac 1{3!}\nabla_{[a}G_{bcd]}
-\frac 14 T_{[ab|}^{~~~E}G_{E|cd]}=0~~~~~~~~~~~~~~~
\eea
We have shown that the super 3-form field strength related to the supergravity covariant derivative algebra (\ref{algebra}) is completely determined in terms of a scalar superfield $G$.  In 3D, a scalar superfield is an irreducible representation of supersymmetry, and therefore the one conventional constraint was enough to completely reduce the super 3-form.


\subsection{Ectoplasmic Integration}
\label{ectoplasmic}

\setcounter{equation}{0}
\renewcommand{\theequation}{\thesubsection.\arabic{equation}}


$~~~$In order to use the Ethereal Conjecture, we must integrate a 3-form over the bosonic sub-manifold.  The super 3-form derived in the previous section is:
\bea
G_{\a\b\g} = 0~~,~~~~~~~~~~~\cr
G_{\a \b c} = (\g_c)_{\a\b}G~~,~~~~~\cr
G_{\a bc} = \e_{bcd}(\g^d)_{\a}^{~\s}\nabla_\s G~~,\cr
G_{abc} = 2\e_{abc}[\nabla^2 G +RG]~~.
\eea
The only problem with this super 3-form is that it has flat indices.  We worked in the tangent space so that we could set supersymmetric constraints on the super 3-form.  Now we require the curved super 3-form to find the generally covariant component 3-form.  In general, the super 3-form with flat indices is related to the super
3-form with curved indices via:\footnote{We have used a different
symbol for the curved super 3-form just to avoid any possible
confusion.}
\bea
\label{flattocurved}
{\cal G}_{MNO}=(-)^{[\frac 32]}{\rm E}{}_M^{~A}{\rm E}{}_N^{~B}
{\rm E}{}_O^{~C}G_{CBA}~~.
\eea
As it turns out, the component 3-form is the lowest component
of the curved super 3-form $g_{mno}={\cal G}_{mno}|$.  Using the usual component definitions for the super frame fields;
E${}_m^{~a}|=e_m^{~a} ,~ {\rm E}_m^{~\a}|=-\psi_{m}^{~\a}$, we can write
the lowest component of the vector 3-form part of (\ref{flattocurved}):
\bea
g_{mno}=-G_{onm}|-\frac 12 \psi_{[m}^{~~\a}G_{no]\a}|
-\frac 12 \psi_{[m}^{~~\a}\psi_{n}^{~~\b}G_{o]\a\b}|
+\psi_m^{~\a}\psi_n^{~\b}\psi_o^{~\g}G_{\a\b\g}|~~.
\eea
Since this is a $\q$ independent equation, we can convert all
of the curved indices to flat ones using e${}_a^{~m}$:
\bea
g_{abc} &=& G_{abc}| -\frac 12 \psi_{[a}^{~\a}G_{bc]\a}|
-\frac 12\psi_{[a}^{~\a}\psi_b^{~\b}G_{c]\a\b}|
+\psi_{a}^{~\a}\psi_{b}^{~\b}\psi_{c}^{~\g}G_{\a\b\g}|\cr
&=&2\e_{abc}[\nabla^2G|+R|G|] -\frac 12 \psi_{[a}^{~\a}
\e_{bc]d}(\g^d)_{\a}^{~\s}\nabla_\s G|-\frac 12
\psi_{[a}^{~\a}\psi_b^{~\b}(\g_{c]})_{\a\b}G| \cr
&=& { \Big \{ } ~ 2\e_{abc}[\, \nabla^2+R| \, ]
-\frac 12 \psi_{[a}^{~\a}\e_{bc]d}(\g^d)_{\a}^{~\s}
\nabla_\s -\frac 12\psi_{[a}^{~\a}\psi_b^{~\b}(\g_{c]}
)_{\a\b} ~ { \Big \}} G|~~.
\eea
We note in passing that this equation is of the form ${\cal D}^2G|$.
Since $g_{abc} $ is part of a closed
super 3-form, it is also closed in the ordinary sense.  Thus,
any volume 3-form $\o^{a b c}$ = $\o$ $\e^{a b c }$ may be
integrated against $g_{abc} $ and will yield an index of the
3D theory if $g_{abc} $ is not exact.  We are led to
define an index $\Delta$ by
\bea
\Delta ~=~ ~ \int \, \o ~ \e^{a b c } \, g_{a b c }~~.
\eea
If we define $\frac 16\e^{abc} g_{abc} $ = ${\cal D}^2G|$ we
can read off the density projector:
\bea
{\cal D}^2 = -2\nabla^2 +\psi_d^{~\a}(\g^d)_\a^{~\s}
\nabla_\s -\frac 12 \psi_a^{~\a}\psi_b^{~\b}\e^{abc}
(\g_c)_{\a\b} -2R~~.
\eea
The Ethereal Conjecture asserts that for all superspace
Lagrangians $\cal L$ the local integration theory for 3D,
$\cal N$ = 1 superspace supergravity takes the form:
\bea
\int d^3 x d^2 \q {\rm E}{}^{-1}{\cal L}=\int d^3 \, {\rm e}{}^{-1}({\cal
D}^2
{\cal L}|)~~.
\eea


\subsection{Obtaining Component Formulations}
\label{component-formalism}

\setcounter{equation}{0}
\renewcommand{\theequation}{\thesubsection.\arabic{equation}}


$~~~$We are interested in describing at the level of component
fields the following general gauge invariant Lagrangian containing two derivatives for 3D, $\cal N$ = 1 gravity
coupled to matter:
\bea \eqalign{
{\cal L} ~=~ & \k^{-2}K(\F)R ~+~ g^{-2}h(\F)_{IJ}W^{\a I} W_\a^J ~+~
g(\F)_{ij}\nabla^\a \F^i \nabla_\a \F^j\cr
&+~ Q_{IJ}\G^I_\b W^{J \b} ~+~ W(\F) ~~~.~~~~~~
} \eea
This action encompasses all possible terms which can arise from the compactification of {\cal M}-theory which we are considering.  The first term is exactly 3D supergravity when $K(\F)=1$.  The second term is the kinetic term for the gauge fields.  The third term is the kinetic part of the sigma model for the scalar matter fields $\F^i$.  The fourth term represents the Chern-Simons term for the gauge fields. Finally, $W(\F)$ is the super potential.

In order to obtain the usual gravity fields we must know how to
define the components of the various field strengths and curvatures.
This is done in a similar manner as before when we determined the
3-form component field of the super 3-form.  In this case, we go to
a Wess-Zumino gauge to write all of the torsions, curvatures and
field strengths at $\q=0$:
\bea
\label{FSX}
T_{ab}^{~~\g}| &=& t_{ab}^{~~\g} ~+~ \psi_{[a}^{~\d}T_{\d b]}^{
~\g} | ~+~ \psi_{[a}^{~\d}\psi_{b]}^{~\g}T_{\d\g}^{~~\g} |~~,\cr
T_{ab}^{~~c}| &=& t_{ab}^{~~c} ~\,+~ \psi_{[a}^{~\d}T_{\d b]}^{
~c} | ~+~ \psi_{[a}^{~\d}\psi_{b]}^{~\g}T_{\d\g}^{~~c} |~~,\cr
R_{ab}^{~~c}| &=& r_{ab}^{~~c}~+~ \psi_{[a}^{~\d}R_{\d b]}^{
~c}| ~+~ \psi_{[a}^{~\d}\psi_{b]}^{~\g}R_{\d\g}^{~~c}  |~~,\cr
{\cal F}_{ab}^{~~I} | &=&  f_{ab}^{~~I} \,+~ \psi_{[a}^{~\d}
{\cal F}_{\d b]}^{~~I}| \,+~ \psi_{[a}^{~\d}\psi_{b]}^{~\g}
{\cal F}_{\d\g}^{~~I} |~~.
\eea
The leading terms in each of these $t_{ab}^{~~\g} $, $t_{ab}^{
~c} $, $r_{ab}^{~~c} $ and $  f_{ab}^{~~I} $ correspond, respectively, to
the exterior derivatives of $\psi {}_a{}^\g$,
e${}_a{}^m$, $\o {}_a{}^c$ and $A{}_a{}^I$, using the bosonic
truncation of the definition of the exterior derivative that
appears in  (\ref{FBianchi1}).  By definition the
super-covariantized curl of the gravitino is the lowest component of the torsion $T_{ab}{}^\g$.  Substituting from (\ref{algebra}) we have:
\bea
f_{ab}^{~~\g}:= T_{ab}^{~~\g}|= -2\e_{abc} \Big[~ \S^{\g c}| +
\fracm 13 (\g^c)^{\g \b} (\nabla_{\b} R)| ~\Big]~~.
\eea
This equation implies the following:
\bea
\label{curlcomp}
\nabla_\a R| = -\frac 14(\g_d)_{\a\g}\e^{abd}f_{ab}^\g~~~,~~~
\S^{\g d}|=\frac 16\Big[\e^{abd}f_{ab}^{~~\g}-(\g_b)_\d^{~\g}
f^{bd\d}\Big]~~.
\eea
The lowest component of $\S^{\d d}$ is indeed gamma traceless.  The
other torsion yields information about the component torsion:
\bea
T_{ab}^{~~c}|=0=t_{ab}^{~~c}+(\g^c)_{\a\b}\psi_{[a}^{~~\a}
\psi_{b]}^{~~\b}~~,
\eea
which can be solved in the usual manner to express the
spin-connection in terms of the anholonomy and gravitino.
The super curvature leads us to the component definition
of the super covariantized curvature tensor:
\bea
R_{ab}^{~~c}|&=&r_{ab}^{~~c}+\psi_{[a}^{~\d}[-2(\g_{b]})_\d^{
~\a}\S_\a^{~c}| -\frac 23(\g_{b]}\g^c)_\d^{~\a}\nabla_\a R|
+\nabla_\d R|\d_{b]}^{~c}] -\psi_{[a}^{~\d}\psi_{b]}^{~\g}(
\g^c)_{\d\g}R|\cr
&=&\e_{abd}[\Hat{\cal R}^{dc}|+\frac 23\h^{cd}(-2\nabla^2R|-\frac
32 R^2|)]~~.
\eea
Contracting this equation with $\e^{ab}{}_c$ and using the component definitions (\ref{curlcomp}) leads to
the component definition:
\bea
\nabla^2R|=-\frac 34R^2| -\frac 14\e^{abc}\psi_a^{~\a}\psi_b^{
~\b}(\g_c)_{\a\b}R| +\frac 18\e^{abc}r_{abc}
+\frac 14\psi^{a\b}(\g^b)_\b^{~\g}f_{ab\g}~~.
\eea
The super field strength satisfies:
\bea
\nonumber
{\cal F}_{ab}^{~~I}| =f_{ab}^{~~I}+\frac 13\psi_{[a}^{~\d}(\g_{b]
})_\d^{~\a}W_\a^I | =\frac 13\e_{abc}(\g^c)_\b^{~\d}\nabla^\b
W_\d^I |~~,
\eea
which implies:
\bea
\nabla_\a W_\b^I| = -\frac 34\e^{abc}(\g_c)_{\a\b}f_{ab}^I -
\frac 12\e^{abc}(\g_c)_{\a\b}(\g_b)_\d^{~\g}\psi_a^{~\d}W_\g^I|~~.
\eea
From (\ref{secondfs}) we have $\nabla_{\a} \nabla^{\b}  W_\b^I$ = 0 so we can derive:
\bea
\nabla^2W_\a^I|=\frac 12(\g^c)_\a^{~\b}\nabla_cW_\b^I|
-\frac 34R|W_\a^I|~~.
\eea
We now have complete component definitions for $R$ and $W_\a$ and enough of the components of $\S^{a\b}$ and $\hat R_{ab}$ to perform the ectoplasmic integration.  Since the gauge potential $\G^I_\a$ for the $U(1)$ fields appears in our Lagrangian we must also make component definitions for it.  $\G^I_\a$ has the gauge transformation:
\bea
\d\G^I_\a=\nabla_\a K^I~~,
\eea
so we can choose the Wess-Zumino gauge:
\bea
\G^I_\a |= \nabla^\a\G^I_\a |= 0~~.
\eea

We are now in a position to derive the full
component action. We introduce the following definitions for the component fields
\bea
R|= B~~~~~W_\a^I|=\l_\a^I~~~~~~~~~~~~~\cr
\F^i|=\f^i~~~~~\nabla_\a\F^i| = \c_\a^i~~~~~\nabla^2\F^i|=F^i \cr
\nabla_\a\G^I_\b|=\frac12(\g^c)_{\a\b}A^I_c~~~~~~~~~\nabla^\b
\nabla_\b\G^I_\a |=\frac 23\l^I_\a~~,
\eea
in addition to the curl of the gravitino defined in
(\ref{FSX}).  Using these component definitions the terms in the action become:
\bea
\label{compact1}
\fracm1{\k^2}\int d^3xd^2\q {\rm E}{}^{-1}K(\F)R ~=\, \,
\fracm1{\k^2}\int d^3x
\, {\rm e}{}^{-1}({\cal D}^2K(\F)R)|~~~~~~~
~~~~~~~~~~~~~~~~~~~~~~~~~~~~~~~~~~\cr
=\, \fracm 1{\k^2}\int d^3x \, {\rm e}{}^{-1}\Big\{
-2B\nabla^2K| +\nabla_\a
K|\Big[-\frac 12 (\g_a)_\b^{~\a}\e^{abc}f_{bc}^{~~\b} -\psi_d^{~\b}
(\g^d)_\b^{~\a}B \Big]
~~~~~~~~~~~~~~~~\cr
+K|\Big[-\frac 12B^2 -\frac 14\e^{abc}
r_{abc} +\frac 14\psi_a^{~\b}\e^{abc}f_{bc\b}\Big]~~
\Big\}~~,~~~
~~~~~\\ \cr
\fracm 1{g^2}\int d^3xd^2\q {\rm E}{}^{-1}h_{IJ}W^{\a I}W_{\a}^{~J}
= \fracm 1{g^2}\int d^3x \, {\rm e}{}^{-1}({\cal D}^2h_{IJ}W^{\a I}W_{\a}^{~J})|
~~~~~~~~~~~~~~~~~~~~~~~~~~~~~\cr
=\fracm 1{g^2}\int d^3 x \, {\rm e}{}^{-1}\Big\{ +\nabla_\a h_{IJ}|\Big[-3
\e^{abc}(\g_c)^{\a\b}f_{ab}^{~~I}\l_\b^{~J} +(\g^a)_\d^{~\a}
\psi_a^{~\d}\l^{\b I}\l_\b^{~J}\Big]
~~~~~~~~~~~~~~~~~~~\cr
+h_{IJ}|\Big[-2(\g^c)^{\a\b}(
\nabla_c\l_\a^{~J})\l_\b^{~J}
-\frac 12\psi^{a\a}\psi_{a\a}\l^{\b I}\l_\b^{~J} +3(\g_e)_\s^{~\r}
f^{deI}\psi_d^{~\s}\l_\r^{~J}\cr
+ B\l^{\b I}\l_\b^{~J} +
\frac 92f^{abI}f_{ab}^{~~J}
-\frac 32\e^{abc}\psi_c^{~\g}\l_\g^{~J}f_{ab}^{~~I}\Big]
-2\nabla^2h_{IJ}|\l^{\b I}\l_\b^{~J}
\Big\}~~,~~~~~\\ \cr
\int d^3x d^2\q {\rm E}{}^{-1}g_{ij}\nabla^\a\F^i\nabla_\a\F^j
= \int d^3 x \, {\rm e}{}^{-1}({\cal D}^2g_{ij}\nabla^\a\F^i\nabla_\a
\F^j)|
~~~~~~~~~~~~~~~~~~~~~~~~~~~~~~~~~~\cr
=\int d^3 x \, {\rm e}{}^{-1}\Big\{
4g_{ij}|\Big[\frac 12(\g^c)_{\a\b}
\nabla_c\c^{\b i} -\frac 14B\c_\a^{~i}\Big]\c^{\a j} + 2
g_{ij}|\Big[-\frac 12\nabla^c\f^i\nabla_c\f^j+2F^iF^j\Big]
~~~~\cr
+g_{ij}|\Big[\psi_d^{~\a}\nabla^d\
f^i\c_\a^{~j} + \e^{abc}(\g_c)_{\b}^{\a}\psi_a^{~\b}\c_\a^i
\nabla_b\f^j + 2(\g^a)_{\a \b}\psi_a^{\b}F^j\c^{\a i}\Big]
~~~~~~~~~\cr
-2\nabla^2g_{ij}|\c^{\a i}\c_\a^{~j}
-\Big[2B+\frac 12\psi_a^{~\a}\psi_b^{
~\b}\e^{abc}(\g_c)_{\a\b}\Big] g_{ij}|\c^{\a i}\c_\a^{~j}
~~~~~~~~~~\cr
+\nabla_\b g_{ij}|\Big[2(\g^c)^{\a\b}\nabla_c\f^i\c_\a^{~j}
-4F^i\c^{\b j}-\psi_d^{~\a}(\g^d)_\a^{~\b}\c^{\a i}\c_\a^{~j}
\Big]
\Big\}~~,~~~~~\\ \cr
\int d^3 x \, {\rm e}{}^{-1}Q_{IJ}\G^I_\b W^{J \b}=\int d^3 x\, {\rm
e}{}^{-1}(
{\cal D}^2Q_{IJ}\G^I_\b W^{J \b})|
~~~~~~~~~~~~~~~~~~~~~~~~~~~~~~~~~
~~~~~~~~~~~~~\cr
=\int d^3 x \, {\rm e}{}^{-1}\Big\{\frac23 Q_{IJ}|\l^I_\b\l^{\b J} -
\nabla^\a Q_{IJ}| (\g^c)_{\a\b}A^I_c\l^{\b J} - \frac12
Q_{IJ}|A^{aI}\psi^{~\a}_a\l^I_\a
~~~~~~~~~~~~~~~~~~\cr
-\frac12 Q_{IJ}|\e^{abc}(\g_a)^{~\b
}_\a A^I_b \psi^{~\a}_c\l^J_\b -\frac32 Q_{IJ}|\e^{abc}
A^I_a f^J_{bc}\Big\}
~~,~~~~~\\ \cr
\label{compact2}
\int d^3 x \, {\rm e}{}^{-1}W(\F)=\int d^3 \, {\rm e}{}^{-1}({\cal
D}^2W)|
~~~~~~~~~~~~~~~~~~~~~~~~~~~~~~~~~
~~~~~~~~~~~~~~~~~~~~~~~~~~~~~~~~~\cr
=\int d^3 x \, {\rm e}{}^{-1}\Big\{ -2\nabla^2W|+\psi_d^{~\a}(\g_d
)_\a^{~\b}\nabla_\b W| - \Big[ 2B
+\frac 12\psi_a^{~\a}\psi_b^{~\b}
\e^{abc}(\g_c)_{\a\b}\Big]W| \Big\}~~.~~~~~~
\eea
This component action is completely off-shell supersymmetric.  We now put it on-shell by integrating out $B$ and $F^i$.  The equation of motion for $F^i$
leads to:
\bea
F^i = \frac 14g^{ij}|\Big\{{\d W\over \d\F^j}\Big| +{\d
g_{kl}\over \d\F^j}\Big|\c^{\a k}\c_\a^{~l} + g^{-2}{\d
h_{IJ}\over \d\F^j}\Big|\l^{\a I}\l_\a^{~J}
~~~~~~~~~~~~~\cr
+ 2\nabla_\a g_{jl}|\c^{\a l}
+\k^{-2}{\d K\over \d\F^j}\Big|B
 -g_{jl}(\g^a)_{\a\b}\psi_a^\b\c^{\a l}\Big\}
~~,~~~
\eea
and the equation of motion for $B$ yields:
\bea
\label{deeznuts}
B=\k^2K|^{-1}\Big\{
g^{-2}h_{IJ}|\l^{\a I}\l_\a^{~J}
-2W| - g_{ij}|\c^{\a i}\c_\a^{~j}
~~~~~~~~~~~\cr
-\k^{-2}(\nabla_\a K|\psi_d^{~\b}(\g^d)_\b^{~\a}
+2\nabla^2K|)
\Big\}~~.
\eea
To be completely general we assume that the coupling
functions depend on some combination of matter fields,
${\cal F}^a$, thus:
\bea
\nabla^2K| &=& \frac 12\sum_a \sum_b{\d^2 K \over \d{\cal
F}^b\d{\cal F}^a}\Big|\nabla^\a{\cal F}^b|\nabla_\a{\cal
F}^a| + \sum_a{\d K\over \d{\cal F}^a}\Big|\nabla^2{\cal
F}^a|\cr
&\equiv&\tilde{\nabla}^2K|+{\d K\over \d \F^i}\Big| F^i~~.
\eea
With this definition, we can substitute for $F^i$ in
(\ref{deeznuts}), leading to:
\bea
\nonumber
B=\k^2K|^{-1}\Big[1+\frac 12\k^{-2}K|^{-1}g^{ij}|{\d K\over\d
\F^i}\Big|{\d K\over\d \F^j}\Big| \Big]^{-1}\Big\{-2W|-\frac
12\k^{-2}g^{ij}|{\d W\over\d \F^i}\Big|{\d K\over\d \F^j}\Big|
\\ \nonumber\cr
-2\k^{-2}\tilde{\nabla}^2K| + \Big[-g_{kl}|-\frac 12\k^{-2}
g^{ij}|{\d K\over\d \F^i}\Big|{\d g_{kl}\over\d \F^j}\Big|
\Big]\c^{\a k}\c_\a^{~l} ~~~~~~~~~~~~~~~~~~\\ \cr
-\k^2g^{ij}|{\d K\over\d \F^i}\Big|\nabla_\a g_{jl}|\c^{\a l}
+\Big[g^{-2}h_{IJ}|-\frac 12\k^{-2}g^{-2}g^{ij}|{\d K\over\d
\F^i}\Big|{\d h_{IJ}\over\d \F^j}\Big| \Big]\l^{\a I}
\l_\a^{~J} \Big\}~~.
\eea
This equation for the scalar field $B$ is what is required to obtain the on-shell supersymmetry variation of the gravitino.  To see this we begin with the off-shell supersymmetry variation of the
gravitino:
\bea
\d_Q \psi_a^{~\b} &=& D_a\e^\b-\e^\a(T_{\a a}^{~~\b}|+T_{\a
a}^{~~b}|\psi_b^{~\b}) - \e^\a\psi_a^{~\g}(T_{\a\g}^{~~\b}|
+T_{\a\g}^{~~e}|\psi_e^{~\b})\cr
&=&D_a\e^\b-\frac 12\e^\a(\g_a)_\a^{~\b}B - \e^\a\psi_a^{~\g}
(\g^e)_{\a\g}\psi_e^{~\b}~~.
\eea
By converting to curved indices and keeping in mind the variation of e${}_m^{~a}$:
\bea
\d_Q {\rm e}{}_{a}{}^{m} = -[\e^\b T_{\b a}^{~~d}|+\e^\b\psi_a^{~\g}
T_{\b\g}^{~~d}|]{\rm e}{}_d^{~m} = -\e^\b\psi_a^{~\g}(\g^d)_{\b\g}
{\rm e}{}_d^{~m}~~,
\eea
the supersymmetry variation of the gravitino can be put into a more canonical form:
\bea
\d_Q \psi_m^{~\b} &=&D_m\e^\b-\frac 12\e^\a(\g_m)_\a^{~\b}B ~~.
\eea
The other fields have the following supersymmetry transformations:
\bea
\d_Q {\rm e}{}_m{}^a &=& \e^\b\phi_m^\g(\g^a)_{\b\g}~~, \cr
\d_Q B &=& \frac14 \e^\a(\g_a)_{\a\g}\e^{abc}f_{bc}^\g~~,\cr
\d_Q A_c^I&=& -\frac13\e^\g(\g_c)_\g^\b\l_\b~~ ,\cr
\d_Q \l_\a^I&=&\e^\b\e^{abc}(\g_c)_{\a\b}(\frac34 f_{ab}^I+\frac12(\g_b)_\d^\g\psi_a^\d \l_\g^I ~~,\cr
\d_Q\f^i&=&-\e^\a\c_\a^i ~~,\cr
\d_Q \c_\a^i&=&-\frac12\e^\b(\g^c)_{\a\b}\nabla_c\f^i + \e_\a  F^i ~~,\cr
\d_Q F^i &=& -\e^\a(\frac12(\g^c)_\a^\b\nabla_c\c_\b^i + \frac14 B\c_\a^i)~~.
\eea
The purely bosonic part of the lagrangian is:
\bea
\label{compbosact}
{\cal S}_{Bosonic}  &=&\int d^3 x \, {\rm e}{}^{-1}\Big[
-\frac12\k^{-2}B^2
- \frac14\k^{-2}\e^{abc} r_{abc} +\frac92 g^{-2} h_{IJ}|f^{ab I}
f_{ab}^J-g_{ij}|\nabla^c\f^i\nabla_c\f^j  \cr
&~& ~~~~~~~~~~~~~~ -\frac32 Q_{IJ}|\e^{abc}A_a^I f_{bc}^J - 2
\partial_i W|F^i -2BW| +4g_{ij}F^i F^j  \Big]~~.
\eea
The equations of motion for B and $F^i$ with $K(\F)$ = 1 and fermions set to zero are:
\bea
\label{ppp-beom}
B=-2\k^2 W| ~~~~~,~~~~~ F^i = \frac14 g^{ij}\partial_j W|~~.
\eea
Substituting these back into the bosonic Lagrangian we have:
\bea
\label{bosons}
{\cal S}_{Bosonic} =\int d^3 x \, {\rm e}{}^{-1}\Big[ -\frac14\k^{-2}
\e^{abc}r_{abc} +\frac92 g^{-2} h_{IJ}|f^{ab I}f_{ab}^J -
g_{ij}|\nabla^c\f^i\nabla_c\f^j
~~~~~~~~~~~\cr
-\frac32 Q_{IJ}|
\e^{abc}A_a^I f_{bc}^J - \, (\, \frac 14 g^{ij}|\partial_i W| \partial_j W|
-2\k^2 W|^2 \, ) ~  \Big]~~.~~~~~
\eea
The scalar potential for this theory can be read off from above
and is given by
\begin{equation}
\label{ppp-sugra-potential}
V(\phi) = \frac 14 g^{ij}\partial_i W| \partial_j W| -2\k^2 W|^2~~,
\end{equation}
and the on-shell supersymmetry variation of the gravitino takes the form
\bea
\label{gavidelta}
\d_Q \psi_m^{~\b} &=&D_m\e^\b  ~-~  \k^2 \, \e^\a (\g_m)_\a^{~\b} \,  W|~~.
\eea
Thus, we see that the superpotential $W$ determines the scalar potential in
the action and appears in the gravitino transformation law. From the form of (\ref{algebra}) and the discussion of the above
section, it is clear the issue of an AdS background is described
in the usual manner known to superspace practicioners. In the limit $R=\sqrt{\lambda}$, ${\Sigma_\alpha}^b=0$, ${W_\alpha}^J=0$ and $\Hat{\cal R}^{ab}=0$ the commutator algebra in (\ref{algebra}) remains consistent in the form
\bea
\label{algebraAdS}
[ \nabla_{\a} ~,~ \nabla_{\b} \} &=&
(\g^c)_{\a \b} ~ \nabla_c ~-~ \sqrt \l \, (\g^c)_{\a \b}{}\, {
\cal M}_c \, ~~~, \cr
~[ \nabla_\a ~,~ \nabla_b \} &=& ~ \frac 12\, \sqrt \l \,
(\g_b)_{\a} {}^{\d}\, \nabla_{\d} ~~,
\cr
[ \nabla_a ~,~ \nabla_b \} &=& - \, \l
\, \e_{abc} \, {\cal
M}^c \, ~~~,
\eea
and clearly the last of these shows that the Riemann curvature tensor
is given by $R_{a \, b}{}^c$ = $- \, \l \, \e_{ab}{}^c $. This
in turn implies that the curvature scalar $\e^{a\, b \, c} \, R_{a \,
b \, c}$ = $- \, 6 \l \, $, i.\ e.\ describes a space of constant
negative curvature. Through the equation for $B$
in (\ref{ppp-beom}) we see that
\be
{\sqrt \l} ~=~ - 2 \, \k^2 \,< W | > \,.
\ee
Thus, there is a supersymmetry preserving AdS backgound whenever the
condition
\be
< W | > ~< ~ 0 \,,
\ee
is satisfied. On the otherhand, supersymmetry is broken whenever
\be
< W | > ~> ~ 0 \,.
\ee
We will see in the next section that the compactified action is of the form (\ref{bosons}).


\section{Compactification of $\cal{M}$-Theory on $Spin(7)$ Holonomy Manifolds}
\label{compactification}

\setcounter{equation}{0}
\renewcommand{\theequation}{\thesection.\arabic{equation}}


$~~~$In this section, we perform the compactification of the
bosonic part of $\cal{M}$-theory on a $Spin(7)$ holonomy manifold
$M_8$.  Since $Spin(7)$ holonomy manifolds admit only one
covariantly constant spinor, we will obtain a theory with $\cn=1$
supersymmetry in three dimensions.  We use the following
assumptions and conventions.  The eight-dimensional manifold
$M_8$ is taken to be compact and smooth\footnote{For an elegant
description of such manifolds see \cite{Joyce}.}. We shall assume
that the size of the internal eight-manifold $l_{M_8} = ({\cal
V}_{M_8})^{1/8}$ is much bigger than the eleven-dimensional
Planck length $l_{11}$. Here ${\cal V}_{M_8}$ denotes the volume
of the internal manifold.

It was shown in \cite{Becker:1996gj, Sethi:1996es, Becker:2000jc}
that compactifications of ${\cal M}$-theory on both conformally
Calabi-Yau four-folds and $Spin(7)$ holonomy manifolds should
obey the tadpole cancellation condition
\begin{equation}
\label{anomaly-condition}
{1 \over 4\kappa^2_{11}} \,
\int_{M_8}\hat{F}^{(1)} \wedge \hat{F}^{(1)} +N_2= T_2 \,
{\chi_8 \over 24}\,,
\end{equation}
where $\hat{F}^{(1)}$ is the internal part of the background
flux, $\chi_8$ is the Euler characteristic of the internal
manifold and $N_2$ represents the number of space-time filling
membranes.  $\kappa_{11}$ is the eleven-dimensional gravitational coupling
constant, which is related to the membrane tension $T_2$ by
\begin{equation}
T_2 = \left({2 \pi^2 \over \kappa_{11}^2}\right)^{1/3} \,.
\end{equation}
Equation (\ref{anomaly-condition}) is important because it
restricts the form of the internal manifold as the Euler
characteristic is expressed in terms of the internal fluxes.  In
our computation, we consider the case $N_2=0$.

In the case when the background fluxes are zero, $\hat F^{(1)}=0$,
the tadpole cancellation condition (\ref{anomaly-condition})
restricts the class of internal manifolds to those which have
zero Euler characteristic. In section \ref{zero-flux}, we
consider this particular case and we show that no scalar
potential for the moduli fields arises under these
circumstances.  To relax the constraint $\chi_8=0$ we have to
consider a non-zero value for the internal background flux
$\hat{F}^{(1)}$. Consequently, we will have to use a warped
metric ansatz. In section \ref{non-zero-flux}, we show that the
appearance of background fluxes generates a scalar potential for
some of the moduli fields appearing in the three-dimensional
${\cal N}=1$ low energy effective action. We also check that the
form of this scalar potential and the complete action is a
particular case of the more general class of models discussed in
the previous section.


\subsection{Compactification with Zero Background Flux}
\label{zero-flux}

\setcounter{equation}{0}
\renewcommand{\theequation}{\thesubsection.\arabic{equation}}


$~~~$The effective action for ${\cal M}\,$-theory has the form
(see e.g.~\cite{Haack:2001jz}, where all relevant references are
provided)
\begin{equation}
\label{start-point}
S=S_0+S_1+S_2\,.
\end{equation}
In this expression $S_0$ is the bosonic truncation of
eleven-dimensional supergravity \cite{Cremmer:1978km} and $S_1$
and $S_2$ represent the leading quantum corrections. The above
terms in the action take the following form
\begin{eqnarray}
\label{zero}
S_0 &=&
{1 \over 2\kappa_{11}^2} \int_{\cal M} d^{11}x\sqrt{-g}R - {1 \over 4\kappa_{11}^2}\int_{\cal M}
\left(F \wedge \star F + {1 \over 3} C \wedge F \wedge F \right)\,,\nonumber \\
\label{pppS_1}
S_1 & = & -T_2 \int_{\cal M} {C \wedge X_8}\,, \nonumber \\
\label{pppS_2}
S_2 & = & b_1 T_2 \int_{\cal M} {d^{11}x \sqrt{-g} \left( J_0 - {1 \over 2}E_8 \right)} \,,
\end{eqnarray}
where $F$ is a four-form field strength with potential $C$ and
$b_1$ is a constant
\begin{equation}
b_1 = {1 \over (2 \pi)^4 3^2 2^{13}} \,.
\end{equation}
$X_8$ is a quartic polynomial of the eleven-dimensional Riemann
tensor, whose integral over the internal manifold is related to the
Euler characteristic
\begin{equation}
X_8({\cal M}) = {1 \over 192 \,(2 \pi)^4} \Big[ {\rm tr} R^4 - {1
\over 4} ({\rm tr} R^2)^2 \Big]\ , \qquad \int_{M_8} X_8\ =\ -
{\chi_8 \over 24} \,.
\end{equation}
Furthermore, $E_8$ and $J_0$ are quartic polynomials in the
eleven-dimensional Riemann tensor, which take the form
\begin{eqnarray}
\label{pppE_8}
E_8({\cal M}) &=& {1 \over 3!} \: \epsilon^{ABCM_1 N_1 \ldots M_4 N_4} \:
\epsilon_{ABCM_1' N_1' \ldots M_4' N_4'} R^{M_1' N_1'} \! _{M_1 N_1} \ldots R^{M_4' N_4'}\! _{M_4
N_4}\,, \\
\label{pppJ_0}
J_0({\cal M}) &=& t^{M_1 N_1 \ldots M_4 N_4} t_{M_1' N_1' \ldots
M_4' N_4'} R^{M_1' N_1'} \! _{M_1 N_1} \ldots R^{M_4' N_4'}\!
_{M_4 N_4} + {1 \over 4} E_8({\cal M})\,.
\end{eqnarray}
The tensor $t$ is defined by its contraction with some
antisymmetric tensor $A$ by
\begin{equation}
t^{M_1 \ldots M_8} A_{M_1 M_2} \ldots A_{M_7 M_8}
  = 24 {\rm tr} A^4 - 6 ({\rm tr} A^2)^2\,,
\end{equation}
and in general we can define $E_n({M_D})$ for any even $n$ and any
$D\,$-dimensional manifold $M_D$ ($n \le D$)
\begin{eqnarray}
E_n(M_D) &=& {1 \over (D-n)!} \: \epsilon^{M_1 \ldots M_{D-n}
N_1 \ldots N_n} \: \epsilon_{M_1 \ldots M_{D-n} N'_1 \ldots
N'_n}\, \nonumber \\
&& R^{N_1' N_2'}\! _{N_1 N_2} \ldots R^{N_{n-1}' N_n'}\! _{N_{n-1}
N_n}\,.
\end{eqnarray}

Our goal is to compactify the action (\ref{start-point}) on a
compact $Spin(7)$ holonomy manifold. In order to achieve this we make
an ansatz for the eleven-dimensional metric
${g}^{(11)}_{MN}(x,y)$, which respects the maximal symmetry of the
external space (described by the metric ${g}^{(3)}_{\mu \nu}(x)$,
which is not necessarily Minkowski)
\begin{equation}
\label{g-ansatz}
{g}^{(11)}_{MN}(x,y) = \left( \begin{array}{cc}
                     {g}^{(3)}_{\mu \nu}(x) & 0 \\
                     0 & {g}^{(8)}_{mn}(x,y)
                     \end{array} \right)\,.
\end{equation}
Here $x$ represents the external coordinates labelled by
$\mu=0,1,2$, while $y$ represents the internal coordinates
labelled by $m=3, \ldots ,10$, and $M, ~N$ run over the complete
eleven-dimensional coordinates.  In addition, ${g}_{mn}(x,y)$
depends on a set of parameters which characterize the possible
deformations of the internal metric. These parameters, called
moduli, appear as massless scalar fields in the three-dimensional
effective action. In other words, an arbitrary vacuum state is
characterized by the vacuum expectation values of these moduli
fields. In the compactification process we choose an arbitrary
vacuum state or equivalently an arbitrary point in moduli space
and consider infinitesimal displacements around this point.
Consequently, the metric will be
\begin{equation}
\label{g-var} {g}_{mn}(x,y) = \hat{g}_{mn}(y) +\delta g_{mn}(x,y)\,,
\end{equation}
where $\hat{g}_{mn}$ is the background metric and $\delta
{g}_{mn}$ is its deformation. The deformations of the metric are
expanded in terms of the zero modes of the Lichnerowicz operator.
Furthermore, it was shown in \cite{Gibbons:1990er}, that for a
$Spin(7)$ holonomy manifold, the zero modes of the Lichnerowicz
operator $e_A$ are in one to one correspondence with the
anti-self-dual harmonic four-forms $\xi_A$ of the internal
manifold
\begin{eqnarray}
\label{e-tensor}
&&e_{A\,mn}(y)={1 \over 6}\,\xi_{A\,mabc}(y) \, {\Omega_{n}}^{abc}(y)\,, \\
\label{e-tensor-inv}
&&\xi_{A\,abcd}(y) = - \,{e_{A\,[a}}^m(y)\Omega_{bcd]m}(y)\,,
\end{eqnarray}
where $A=1, \ldots b_4^-$ and $\Omega$ is the Cayley calibration
of the internal manifold, which in our convention is
self-dual.  The tensor $e^I_{mn}$ is symmetric and
traceless (see \cite{Gibbons:1990er}). $b_4^-$ is the Betti
number that counts the number of anti-self-dual harmonic
four-forms of the internal space.

Besides the zero modes of the Lichnerowicz operator there is an
additional volume-changing modulus, which corresponds to an
overall rescaling of the background metric. So the metric
deformations take the following form
\begin{equation}
\label{g-expan}
\delta g_{mn}(x,y)=  \phi(x) \, \hat{g}_{mn}(y) + \sum_{A=1}^{b_4^-} \phi^A(x) \, e_{A\,mn}(y) \,,
\end{equation}
where $ \phi$ is the radial modulus fluctuation and $\phi^A$ are
the scalar field fluctuations that characterize the deformations
of the metric along the directions $e_A$.  Therefore the internal
metric has the following expression
\begin{equation}
\label{internal-metric} g_{mn}(x,y)= \hat{g}_{mn}(y) + \phi(x) \,
\hat{g}_{mn}(y) + \sum_{A=1}^{b_4^-}\phi^A(x) \, e_{A\,mn}(y) \,.
\end{equation}

The three-form potential and the corresponding field strength have
fluctuations around their backgrounds $\hat C(y)$ and $\hat F(y)$
respectively, which in this section are considered to be zero.
The fluctuations of the three-form potential are decomposed in
terms of the zero modes of the Laplace operator. Taking into
account that for $Spin(7)$ holonomy manifolds there are no
harmonic one-forms (see (\ref{cohomology}) ) the decomposition of
the three-form potential has two pieces
\begin{equation}
\label{dc-expand} \delta C(x,y) = \delta C^{(1)}(x,y) + \delta
C^{(2)}(x,y) = \sum_{I=1}^{b_2}  A^I(x) \wedge \omega_I(y) +
\sum_{J=1}^{b_3} \rho^J(x) \, \zeta_J(y)\,,
\end{equation}
where $\omega_I$ are harmonic two-forms and $\zeta_J$ are harmonic
three-forms. The set of $b_2$ vector fields $A^I(x)$ and the set
of $b_3$ scalar fields $\rho^J(x)$ are infinitesimal quantities
that characterize the fluctuation of the three-form potential
around its background value. The fluctuations of the field
strength $F$ are then
\begin{equation}
\label{df-expand} \delta F(x,y) = \delta F^{(1)}(x,y) + \delta
F^{(2)}(x,y) = \sum_{I=1}^{b_2} dA^I(x) \wedge \omega_I(y) +
\sum_{J=1}^{b_3} d\rho^J(x) \wedge \zeta_J(y)\,.
\end{equation}

Substituting (\ref{internal-metric}), (\ref{dc-expand}) and
(\ref{df-expand}) into $S$ and considering the lowest order
contribution in moduli fields we obtain
\begin{eqnarray}
\label{zero-flux-action}
S_{3D}&=& {1 \over 2\kappa_3^2} \,
\int_{M_3}d^3x \, \sqrt{-g^{(3)}} \, \Big\{ R^{(3)} - 18
(\partial_\alpha\phi)(\partial^\alpha\phi)  - \sum_{A,B=1}^{b_4^-}
{\cal L}_{AB} (\partial_\alpha \phi^A)
(\partial^\alpha \phi^B) \nonumber \\
&& - \sum_{I,J=1}^{b_3} {\cal M}_{IJ}  (\partial_\alpha\rho^I) (\partial^\alpha\rho^J) - \sum_{I,J=1}^{b_2} {\cal K}_{IJ} ~ f_{\alpha\beta}^I
 f^{J \alpha \beta} \Big\}~+~\ldots \,,
\end{eqnarray}
where the ellipses denote higher order terms in moduli
fluctuations. $\kappa_3$ is the three-dimensional gravitational
coupling constant
\begin{equation}
\kappa_3^2 = {\cal V}^{-1}_{M_8} \, \kappa_{11}^2 \,,
\end{equation}
and ${\cal V}_{M_8}$ is the volume of the internal manifold
\begin{equation}
\label{volume}
{\cal V}_{M_8} = \int_{M_8} d^8y \, \sqrt{\hat{g}^{(8)}} \,.
\end{equation}
The details of the dimensional reduction of the Einstein-Hilbert
term can be found in appendix \ref{appendix-zero-flux}. The other
quantities appearing in (\ref{zero-flux-action}) are the field
strength $f^I$ of the $b_2$ $U(1)$ gauge fields $A^I$
\begin{equation}
f^I_{\alpha \beta} = \partial_{[\alpha}A^I_{\beta]} = {1 \over 2}
(\partial_\alpha A^I_\beta-\partial_\beta A^I_\alpha) \,,
\end{equation}
and the metric coefficients for the kinetic terms
\begin{eqnarray}
\label{moduli-metric-l}
{\cal L}_{AB} &=& {1 \over 4{\cal V}_{M_8}}\int_{M_8}d^8y\,\sqrt{\hat{g}^{(8)}} \, e_{A \, am} \,
e_{B\,bn} \, \hat{g}^{ab}\, \hat{g}^{mn}\,,\\%
\label{moduli-metric-k}
{\cal K}_{IJ} &=& {3 \over 2 {\cal V}_{M_8}} \, \int_{M_8} \omega_I \wedge
\star \, \omega_J \,,\\%
\label{moduli-metric-m} {\cal M}_{IJ} &=& {2 \over {\cal V}_{M_8}} \,
\int_{M_8} \zeta_I \wedge \star \, \zeta_J \,.
\end{eqnarray}
With the help of (\ref{e-tensor}) and (\ref{e-tensor-inv}) we can rewrite
(\ref{moduli-metric-l}) as follows
\begin{equation}
\label{moduli-metric-l2}
{\cal L}_{AB} = {1 \over {\cal V}_{M_8}} \, \int \xi_A \wedge \star \, \xi_B \,.
\end{equation}
Note that the Hodge $\star$ operator used in the previous
relations is defined with respect to the background metric. As we
can see in the zero flux case, the action contains only the
gravitational part plus kinetic terms of the massless moduli
fields and no scalar potential.


\subsection{Compactification with Non-Zero Background Flux}
\label{non-zero-flux}

\setcounter{equation}{0}
\renewcommand{\theequation}{\thesubsection.\arabic{equation}}


$~~~$Our goal in this section is to compute the form of the three-dimensional
effective action in the presence of non-vanishing fluxes. We begin by decomposing
the metric and flux fields and then work out the compactification.  We make the
following maximally symmetric ansatz for the metric\\
\begin{equation}
\label{warped-metric}
\widetilde{g}^{(11)}_{MN}(x,y) = \left( \begin{array}{cc}
                     e^{2\Delta(y)/3} \, {g}^{(3)}_{\mu \nu}(x) & 0 \\
                     0 & e^{-\Delta(y)/3} \, {g}^{(8)}_{mn}(x,y)
                     \end{array} \right)\,,
\end{equation}
where $\Delta(y)$ is the scalar warp factor, ${g}^{(3)}_{\mu
\nu}(x)$ is the metric for the external space and
${g}^{(8)}_{mn}(x,y)$ has $Spin(7)$ holonomy. Maximal symmetry of
the external space restricts the form of the background flux to
\begin{eqnarray}
\label{bf-expand}
\hat{F}(y) &=& \hat{F}^{(1)}(y)+\hat{F}^{(2)}(y)\,,\nonumber \\
\hat{F}^{(1)}(y) &=& {1 \over 4!} \, F_{mnpq}(y) \: dy^m \wedge dy^n \wedge dy^p
\wedge dy^q\,, \nonumber \\\hat{F}^{(2)}(y) &=& {1 \over 3!} \, \epsilon_{\alpha
\beta\gamma}
\partial_m f(y) \: dx^\alpha \wedge dx^\beta \wedge dx^\gamma
\wedge dy^m\,,
\end{eqnarray}
therefore, $C$ has the following background
\begin{eqnarray}
\label{bc-expand}
\hat{C}(y) &=& \hat{C}^{(1)}(y)+\hat{C}^{(2)}(y)\,,\nonumber \\
\hat{C}^{(1)}(y) &=& {1 \over 3!} \, C_{mnp}(y) \: dy^m \wedge dy^n \wedge
dy^p\,, \nonumber \\\hat{C}^{(2)}(y) &=& -\,{1 \over 3!} \, \epsilon_{\alpha
\beta\gamma} f(y) \: dx^\alpha \wedge dx^\beta \wedge dx^\gamma\,.
\end{eqnarray}
In addition, $f(y)$ is related to the warp factor $\Delta(y)$ by \cite{Becker:2000jc}
\begin{equation}
f(y)=e^{\Delta(y)}\,.
\end{equation}
The 5-brane Bianchi identity derived in \cite{Becker:2000jc}
implies that the warp factor is a small quantity, $\Delta \sim
{\cal O}(l_{11}^6/l_8^6)\,$.  Further, the tadpole anomaly
equation (\ref{anomaly-condition}) implies that the internal flux
is also small, $\hat{F}^{(1)} \sim {\cal O}(l_{11}^3/l_8^4)\,$.
Consequently, we will consider only the leading order
contribution.

Next we start the compactification of the eleven-dimensional
action. The Einstein-Hilbert term becomes
\begin{eqnarray}
\label{warped Einstein-Hilbert}
{1 \over 2\kappa_{11}^2}\,\int_{\cal M} \, d^{11}x \,\sqrt{-\widetilde{g}^{(11)}}\,
\widetilde{R}^{(11)} \, &=& \, {1 \over 2\kappa_3^2}\,\int_{M_3}
d^3x \,\sqrt{-g^{(3)}}\,  \Big\{ R^{(3)} - 18(\partial_\alpha \phi) \, (\partial^\alpha \phi)
\nonumber\\&&-\, \sum_{I,J=1}^{b_4^-} \, {\cal L}_{IJ} \,(\partial_\alpha
\phi^I) \, (\partial^\alpha \phi^J)\Big\} ~+~\ldots \,,
\end{eqnarray}
where ${\cal L}_{IJ}$ was defined in section \ref{zero-flux}. The
details of the dimensional reduction can be found in appendix
\ref{appendix-non-zero-flux}.  The quartic polynomial $J_0$,
(\ref{pppJ_0}), which appears in the definition (\ref{pppS_2}) of
$S_2$, is the sum of an internal and external polynomial.
Further, it can be written in terms of the Weyl tensor
\cite{Banks:1998nr, Gubser:1998nz}. Since the Weyl tensor
vanishes in three dimensions we are left with only the internal
polynomial $J_0(M_8)$. We shall restrict to compactifications on
manifolds with $J_0(M_8)=0$. It is rather possible that this
holds for a generic $Spin(7)$ holonomy manifold, however this
needs to be evaluated in more detail \cite{Dragos}. The final
piece of $S_2$ involves the quartic polynomial $E_8$.  For a
product space $M_3 \times M_8$ we have \cite{Haack:2001jz}
\begin{equation}
\label{e8-decomp}
E_8(M_3 \times M_8) = -E_8(M_8) + 4 \,E_2(M_3) \, E_6(M_8) \,,
\end{equation}
where $E_2(M_3)=-2R^{(3)}\,$. Therefore
\begin{equation}
b_1 T_2 \int_{\cal M} {d^{11}x \,\sqrt{-g}\, \left( J_0 - {1 \over 2}E_8
\right)}  = \int_{M_3}  d^3x \,\sqrt{-g^{(3)}}\,  T_2 \,
{\chi_8 \over 24}~+~\dots \,,
\end{equation}
where we have used the fact that the Euler characteristic of the
internal manifold is
\begin{equation}
\chi_8 = 12b_1 \, \int_{M_8} d^8y\,\sqrt{\hat{g}^{(8)}} \, E_8(M_8)\,,
\end{equation}
and we have neglected the subleading contribution from the second
term of (\ref{e8-decomp}).  This concludes the analysis of the
terms in $S$ that only depend on the metric.

The remaining terms in $S$ consist of the kinetic term for $C$,
the Chern-Simons term, and the tadpole anomaly term $S_1$. The
expressions (\ref{df-expand}) and (\ref{bf-expand}) of the field
strength $F$ imply that
\begin{eqnarray}
\int_{\cal{M}} F \wedge \star \,F &=&\int_{\cal{M}} \hat{F}^{(1)}
\wedge \star \, \hat{F}^{(1)} + \int_{\cal{M}} \hat{F}^{(2)}
\wedge \star \, \hat{F}^{(2)} \nonumber \\
&+& \int_{\cal{M}} \delta F^{(1)} \wedge \star \, \delta F^{(1)} +
\int_{\cal{M}} \delta F^{(2)} \wedge \star \, \delta F^{(2)}\,,
\end{eqnarray}
where the second term is subleading and will be neglected. To
leading order, the last two terms in the above sum can be
expressed as
\begin{eqnarray}
&&{1 \over 4\kappa_{11}^2}\,\int_{\cal{M}}[\delta F^{(1)} \wedge
\star \, \delta F^{(1)} + \delta F^{(2)} \wedge \star \, \delta F^{(2)}] \nonumber\\
&& = {1 \over 2\kappa_3^2}\,\int_{M_3}d^3x \, \sqrt{-g^{(3)}} \, \Big\{
\sum_{I,J=1}^{b_3} {\cal M}_{IJ} \, (\partial_\alpha\rho^I)
(\partial^\alpha\rho^J) \nonumber\\
&&+ \sum_{I,J=1}^{b_2} {\cal K}_{IJ} \, f_{\alpha\beta}^I\: f^{J \alpha
\beta}\Big\},
\end{eqnarray}
where $f^I$, ${\cal K}_{IJ}$ and ${\cal M}_{IJ}$ were defined in
section \ref{zero-flux}.  Due to specific structure of $C(x,y)$
and $F(x,y)$ given in equations (\ref{dc-expand}),
(\ref{df-expand}), (\ref{bf-expand}) and (\ref{bc-expand}) the
Chern-Simons term will have the following form to leading order
in moduli field fluctuations
\begin{equation}
\label{chern-simons}
\int_{\cal{M}}C \wedge F \wedge F  = 3
\int_{\cal{M}} \hat{C}^{(2)} \wedge \hat{F}^{(1)} \wedge
\hat{F}^{(1)} + 2 \int_{\cal{M}} \delta C^{(1)} \wedge \delta F^{(1)} \wedge
\hat{F}^{(1)} ~+~\ldots \,.
\end{equation}
Since the first term in (\ref{chern-simons}) cancels the tadpole anomaly
term, $S_1$, the sum of the Chern-Simons term and $S_1$ is
\begin{equation}
\label{chern-simons-final}
{1 \over 12 \kappa_{11}^2} \, \int_{\cal{M}}C \wedge F \wedge F +
T_2 \int_{\cal M} {C \wedge X_8} = {1 \over 6 \kappa_{11}^2} \, \int_{\cal{M}} \delta C^{(1)}
\wedge\delta F^{(1)} \wedge \hat{F}^{(1)} ~+~\ldots \,.
\end{equation}
Using (\ref{dc-expand}) and (\ref{df-expand}) we obtain
\begin{equation}
{1 \over 6 \kappa_{11}^2} \, \int_{\cal{M}} \delta C^{(1)}
\wedge\delta F^{(1)} \wedge \hat{F}^{(1)} = {1 \over 2\kappa_3^2}\,\sum_{I,J=1}^{b_2}
{\cal E}_{IJ}\int_{M_3} A^I \wedge dA^J\,, \end{equation}
where we have defined
\begin{equation}
\label{cs-coef}
{\cal E}_{IJ}={1 \over 3 {\cal V}_{M_8}} \, \int_{M_8}\omega^I \wedge \omega^J \wedge
\hat{F}^{(1)}\,.
\end{equation}
The coefficient (\ref{cs-coef}) is proportional to the internal
flux and this is the reason why we did not obtain such a term in
section \ref{zero-flux}. We have completed the compactification
of $S$.  Using the above formulas we obtain to leading order in
moduli fields the following expression for the low energy
effective action
\begin{eqnarray}
\label{non-zero-flux-action}
S_{3D} &=& {1 \over 2\kappa_3^2} \,
\int_{M_3}d^3x \, \sqrt{-g^{(3)}} \, \Big\{ R^{(3)} - 18
(\partial_\alpha\phi)(\partial^\alpha\phi)  - \sum_{I,J=1}^{b_4^-}
{\cal L}_{IJ} (\partial_\alpha \phi^I)
(\partial^\alpha \phi^J) \nonumber \\
&& - \sum_{I,J=1}^{b_3} {\cal M}_{IJ} \, (\partial_\alpha\rho^I) (\partial^\alpha\rho^J) - \sum_{I,J=1}^{b_2} {\cal K}_{IJ} \, f_{\alpha\beta}^I
\: f^{J \alpha \beta} \nonumber\\
&&- \sum_{I,J=1}^{b_2} {\cal E}_{IJ} ~ \epsilon^{\mu \nu \sigma}
A_\mu^I ~ f_{\nu \sigma}^J - V \Big\} + \ldots \,,
\end{eqnarray}
where the scalar potential $V$ is
\begin{equation}
\label{potential}
V =  \, {1 \over 2{\cal V}_{M_8}} \,  \int_{M_8}\hat{F}^{(1)}\wedge
\star \hat{F}^{(1)} - 2 \kappa_3^2~T_2 \, {\chi_8  \over 24} \,.
\end{equation}
This potential is a particular case of the more general
construction presented in the previous section
(\ref{ppp-sugra-potential}). Let us elaborate this in detail.

First, we will show that the scalar potential (\ref{potential})
can be written in terms of the superpotential
\begin{equation}
\label{ppp-superpotential} W=\int_{M_8}\hat{F}^{(1)} \wedge
\Omega \,.
\end{equation}
The form of the superpotential $W$ was conjectured in
\cite{Gukov:1999gr}. This conjecture has been checked recently in
\cite{Acharya:2002vs, Becker:2002jj}. More explicitly in
\cite{Becker:2002jj} the supersymmetry transformation for the
gravitino (\ref{gavidelta}) was used to obtain the form of the
superpotential. Using the anomaly cancellation condition
(\ref{anomaly-condition}), the scalar potential becomes
\begin{equation}
V = \, {1 \over {\cal V}_{M_8}} \, \int_{M_8} \hat{F}^{(1)}_-
\wedge \star \, \hat{F}^{(1)}_-\ \,,
\end{equation}
where
\begin{equation}
\hat{F}^{(1)}_- = {1 \over 2} \left[ \hat{F}^{(1)} - \star \, \hat{F}^{(1)} \right] \,,
\end{equation}
is the anti-self-dual part of the internal flux $\hat{F}^{(1)}$.
Using the definition (\ref{moduli-metric-l2}) for ${\cal L}_{AB}$
we can obtain the functional dependence of the scalar potential
$V$ in terms of the superpotential (\ref{ppp-superpotential})
\begin{equation}
\label{scalar-potential} V[W] = \sum_{A,B=1}^{b_4^-} {\cal L}^{AB} \, D_AW \, D_BW \,,
\end{equation}
where ${\cal L}^{AB}$ is the inverse matrix of ${\cal L}_{AB}$ and we have introduced the operator
\begin{equation}
\label{D-operator}
D_A \Omega = \partial_A \Omega + K_A \Omega \,.
\end{equation}
As shown in the appendix \ref{appendix-spin7} (see
\ref{partail-a-omega}) if the $D_A$ operator acts on the Cayley
calibration the result is an anti-self-dual harmonic four-form.

What we observe from (\ref{scalar-potential}) is that the
external space is restricted to three-dimensional Minkowski
because the scalar potential is a perfect square, in agreement
with \cite{Acharya:2002vs}. Furthermore, when $D_A W=0$ the scalar
potential vanishes. This gives us a set of $b_4^-$ equations for
$b_4^- + 1$ fields, so that the radial modulus is not fixed at
this level. Its rather possible that non-perturbative effects
will lead to a stabilization of this field, as in
\cite{Kachru:2003aw}.

A few remarks are in order before we can compare the compactified
action to the supergravity action.  For a consistent analysis, we
must take into account all of the kinetic terms for the metric
moduli.  Furthermore, the scalar potential
(\ref{scalar-potential}) does not seem to be a special case of
(\ref{ppp-sugra-potential}). The discrepancy arises for two
reasons.  First, in the general case the superpotential may
depend on all of the scalar fields existing in the theory and the
summation in (\ref{ppp-sugra-potential}) is taking into account
all of these scalars, whereas in the compactified version the
superpotential depends only on the metric moduli
\begin{equation}
\partial_i W \neq 0 ~~~~~ i=0,~1, \ldots b_4^- \,,
\end{equation}
where $``0"$ labels the radial modulus. Second, as described in
(\ref{partial-0-omega}) the superpotential has a very special
radial modulus dependence in the sense that
\begin{equation}
\partial_0 W = 2W \,,
\end{equation}
and this is the reason why the summation in
(\ref{scalar-potential}) does not include the radial modulus.
Keeping these remarks in mind we proceed to show that the result
coming from compactification is a particular case of the general
supergravity analysis.

We begin by rescaling some of the fields in the supergravity
action
\begin{eqnarray}
2 \kappa_3^2~g_{ij}| &=& L_{ij} \,, \nonumber \\
{1 \over 2 \kappa_3^2}~g^{ij}| &=& L^{ij} \,, \\
\kappa_3^2~ W| &=& \widetilde{W} \,. \nonumber
\end{eqnarray}
The relevant terms in the supergravity action (\ref{bosons}) can be written as
\begin{eqnarray}
\int_{M_3}d^3x ~ \sqrt{-g^{(3)}} ~ \Big\{  - g_{ij} ~
\partial_\alpha \bar{\f}^i ~ \partial^\alpha \bar{\f}^j  -
[\, \frac 14 g^{ij} ~ \bar{\partial}_i W ~ \bar{\partial}_j W
- 2\k_3^2 ~ W^2 \, ] ~  \Big\} \nonumber
~~~~~~~~~~\\
={1 \over 2 \kappa_3^2}~\int_{M_3}d^3x ~ \sqrt{-g^{(3)}} ~
\Big\{ - L_{ij} ~ \partial_\alpha \bar{\f}^i ~ \partial^\alpha \bar{\f}^j
\label{qqq-sugra}
- [\, L^{ij} ~ \bar{\partial}_i \widetilde{W} ~ \bar{\partial}_j \widetilde{W}
- 4 ~ \widetilde{W}^2 \, ] ~  \Big\} \,.
\end{eqnarray}
In the above equation the indices $i,~j = 0,~ 1, \ldots b_4^-$. In
what follows we will drop the label ``0'' from the radion
$\phi_0=\phi$ and the derivative with respect to it $\partial_0=\partial$ and we will
denote with $A,~B \ldots$ the remaining set of indices, i.e.
$A,~B=1,\ldots,b_4^-$. We have placed bars on the scalar fields
and their derivatives in (\ref{qqq-sugra}) to avoid confusion
since we require one more field redefinition.

The relevant terms from the compactified action have
the following form
\begin{eqnarray}
{1 \over 2\kappa_3^2} \int_{M_3}d^3x \sqrt{-g^{(3)}} \Big\{ -18 (\partial \phi)^2 -{\cal L}_{AB}
(\partial_\alpha\phi^A)(\partial^\alpha\phi^B)  - {\cal L}^{AB} D_A W D_B W ~\Big\} \nonumber \\
=  \int_{M_3}d^3x \, \sqrt{-g^{(3)}} \, \Big\{ -18 (\partial_{\alpha} \phi)(\partial^{\alpha}\phi)
-{\cal L}_{AB} (\partial_\alpha\phi^A)(\partial^\alpha\phi^B) \nonumber
~~~~~~~~~~~~~~\\
\label{qqq-compact}
 - [{\cal L} ^{AB} (\partial_A W) (\partial_B W)
+ 4 {\cal L}^A (\partial_A W)W + {\cal L} W^{2}] ~\Big\}.
\end{eqnarray}
In the above equation we have used the expression
(\ref{D-operator}) for $D_A$ and we have introduced ${\cal
L}^{A}={\cal L}^{AB}K_B$ and ${\cal L}={\cal L}^{AB}K_AK_B$. In
order to make the comparison between (\ref{qqq-sugra}) and
(\ref{qqq-compact}), we have to redefine the fields in
(\ref{qqq-sugra}) in the following manner
\begin{eqnarray}
&&\phi = L_{00} \bar{\phi} + L_{0A} \bar{\phi^A} \,, \nonumber \\
&&\phi^A = \bar{\phi^A} \,.
\end{eqnarray}
Keeping track that $\phi$ is the radial modulus, we obtain the
following form for the relevant terms of the supergravity action
\begin{eqnarray}
{1 \over 2\kappa_3^2} \int_{M_3}d^3x \sqrt{-g^{(3)}} \Big\{ -L_{ij}(\partial_\alpha \bar{\phi}^i)
(\partial^\alpha \bar{\phi}^j)  - [ L^{ij}(\bar{\partial_i} \widetilde{W})
(\bar{\partial_j} \widetilde{W}) -4 \widetilde{W}^2 ] ~\Big\} \nonumber
~~~~~~~~~~\\
= {1 \over 2\kappa_3^2} \int_{M_3}d^3x \sqrt{-g^{(3)}} \Big\{ - {1 \over L_{00}}
(\partial_\alpha \phi)(\partial^\alpha \phi) - ( L_{AB}-{L_{0A} L_{0B} \over L_{00}} )
(\partial_\alpha \phi^A)(\partial^\alpha \phi^B) \nonumber \\
- [ L^{AB} (\partial_A \widetilde{W})(\partial_B \widetilde{W})
+ 4L^{0A} L_{00} \widetilde{W} (\partial_A \widetilde{W}) + 4(L^{00}L_{00}^2+3L_{00}+1) \widetilde{W}^2]\Big \} \,.
\end{eqnarray}
Surprisingly, we have that
\begin{equation}
(L_{AB}-{L_{0A}L_{0B} \over L_{00}})L^{BC}=\delta_A^C,
\end{equation}
and as a consequence we can perform the following identifications
\begin{eqnarray}
&&L_{00}={1 \over 18} \,, \nonumber \\
&&L_{AB}-{L_{0A}L_{0B} \over L_{00}} = {\cal L}_{AB} \,, \nonumber \\
&&L^{AB}={\cal L}^{AB} \,, \nonumber \\
&&L^{0A}L_{00} = {\cal L}^{AB}K_B \,, \nonumber \\
&& 4(L^{00}L_{00}^2+3 L_{00}+1) ={\cal L}^{AB}K_A K_B \,.
\end{eqnarray}
With these identifications, both actions are seen to coincide.
The remaining kinetic terms and the Chern-Simons terms that were
left in the actions (\ref{bosons}) and
(\ref{non-zero-flux-action}) can be easily identified and we
conclude that the compactified action is in perfect agreement with
the general supergravity action. ${\cal M}$-theory compactified on
manifolds with $Spin(7)$ holonomy produces a low energy effective
action that corresponds to a particular case of the minimal three
dimensional supergravity coupled with matter.


\section{Summary and Open Questions}
\label{open-questions}

\setcounter{equation}{0}
\renewcommand{\theequation}{\thesection.\arabic{equation}}


$~~~$In this paper we have derived the general form of 3D, ${\cal
N}=1$ supergravity coupled to matter.  The off-shell component
action is the sum of (\ref{compact1}-\ref{compact2}), the on-shell
bosonic action is given in (\ref{bosons}), and the supersymmetry
variation of the gravitino, (\ref{gavidelta}), was shown to be
proportional to the superpotential. The latter statement was an
important ingredient in order to check \cite{Becker:2002jj} the
form of the superpotential for compactifications of ${\cal
M}$-theory on $Spin(7)$ holonomy manifolds conjectured in
\cite{Gukov:1999gr}. We have also performed the Kaluza-Klein
compactification of ${\cal M}$-theory on a $Spin(7)$ holonomy
manifold with and without fluxes.  When fluxes are included, we
generate a scalar potential for moduli fields. This scalar
potential can be expressed in terms of the superpotential,
(\ref{ppp-superpotential}). Interestingly, the potential
(\ref{scalar-potential}) is a perfect square, so that only
compactifications to three-dimensional Minkowski space can be
obtained in agreement with \cite{Acharya:2002vs}. It is plausible
that non-perturbative effects will modify this result to
three-dimensional de-Sitter space along the lines of
\cite{Kachru:2003aw}. This will be an interesting question for
the future.

Another interesting issue we addressed is the duality between a
strongly coupled gauge theory and a weakly coupled supergravity
theory. Recall that the supergravity dual to the 4D confining
gauge theory given by Polchinski and Strassler
\cite{Polchinski:2000uf} has yet to be found to all orders in
perturbation theory.  This verification could be obtained by
considering a generalization of the compactifications of ${\cal
M}$-theory on eight manifolds of \cite{Becker:1996gj}. Work in
this direction has been done recently in \cite{Martelli:2003ki,
Frey:2003sd, Becker:2003xd}, though the complete supergravity
dual to all orders is still lacking. In a similar vein, it would
be interesting to find gauge theories which are dual to
compactifications of ${\cal M}$-theory on $Spin(7)$ holonomy
manifolds. These theories would be 3D conformal gauge theories
with ${\cal N}=1$ supersymmetry. It may then be possible to
deform the $Spin(7)$-manifold to obtain a confining gauge theory.


\bigskip
\bigskip
\noindent {\Large\bf Acknowledgments}

We would like to thank Katrin Becker, Sergei Gukov, Michael
Haack, Dominic Joyce, Axel Krause, Peter van Nieuwenhuizen, Joseph Polchinski,
Martin Ro\v{c}ek, Ram Sriharsha, Arkady Tseytlin and Edward Witten for useful
discussions. This work is supported by the NSF under grant
PHY-0244722 and the Center for String and Particle
Theory at the University of Maryland. The work of M. Becker was
partially supported by an Alfred Sloan Fellowship. J. Phillips would like
to thank the warm hospitality of Pierre Ramond and the Institute for Fundamental
Theory at the University of Florida, Gainesville, where part of
this work has been done.

\pagebreak


\appendix


\section{Notations and Conventions}
\label{conventions}


\subsection{3D Supergravity}
\label{3d-sugra-conventions}

\setcounter{equation}{0}
\renewcommand{\theequation}{\thesubsection.\arabic{equation}}


$~~~$We use lower case Latin letters for three-vector indices and
Greek letters for spinor indices.  Supervector indices are denoted
by capital Latin letters $A, M$.  We further employ the early late
convention: letters at the beginning of the alphabet denote
tangent space indices while letters from the middle of the
alphabet denote coordinate indices. The spinor metric is defined
through:
\bea
C_{\m \, \n} C^{\s \, \t} ~=~ \d_{\m}{}^{\s} \, \d_{\n}{}^{\t}
~-~  \d_{\m}{}^{\t} \, \d_{\n}{}^{\s} ~ \equiv ~
\d_{\m}{}^{[ \s} \, \d_{\n}{}^{\t ]}
~~~,
\eea
and is used to raise and lower spinor indices via:
\bea
\q_{\n} ~=~ \q^{\m} \, C_{\m \, \n}  ~~~,~~~
\q^{\m} ~=~  C^{\m \, \n}  \, \q_{\n}
~~~.
\eea
Some other conventions:
\bea
diag (\, \eta_{ab} \, ) ~=~ (-1 , 1 , 1)  ~~~,~~~ \e_{a b c} \, \e^{d e f} ~=~
-\d_{[a}{}^d \d_{b}{}^e \d_{c]}{}^f ,~~~ \e^{0 1 2} =  +1 ~~~.
\eea
The $\g$-matrices are defined through:
\bea
\label{gamma}
(\g^a)_{\a}{}^{ \g} (\g^b)_{\g}{}^{ \b} =
\eta^{ ab} \d_{\a}{}^{\b} +  \e^{ a
b c}  (\g_c)_{\a}{}^{\b} ~~,
\eea
and satisfy the Fierz identities:
\bea
(\g^a)_{\a \b} (\g_a)^{\g \d} ~=~ - \d_{(\a}{}^\g
\d_{\b)}{}^\d ~=~ - (\g^a)_{(\a}{}^\g (\g_a)_{\b)}{}^\d  ~~, ~~~~~~~\\
\e^{a b c} \, (\g_b)_{\a \b} (\g_c)_\g {}^\d ~=~ C_{\a \g} (\g^a)_\b
{}^\d +(\g^a)_{\a \g} \d_{\b} {}^\d ~~~.~~~~~
\eea
For the Levi-Cevita symbol, we have the contractions:
\bea
\e^{abc}\e_{def}=-\d_{[d}^{~~a}\d_{e}^{~b}\d_{f]}^{~c}~~,\cr
\e^{abc}\e_{dec}=-\d_{[d}^{~~a}\d_{e]}^{~b}~~,\cr
\e^{abc}\e_{dbc}=-2\d_{d}^{~a}~~.
\eea
The Lorentz rotation generator is realized in the following manner:
\bea
e^{-\frac 12 \l_{ab}{\cal M}^{ab}}=
e^{-\frac 12 \e_{abc} \l^c \frac 14 \g^{[a}\g^{b]}}=
e^{+\frac 12 \l^c \g_c}~~.
\eea
Infinitesimally, the action of the Lorentz generator is:
\bea
[~ {\cal M}_a  \, , \,\varphi(x) \,
~ ] &=&~ 0 ~~~, \cr
 [~{\cal M}_a \, ,
\, \r_{\a} (x)
\,  ] &=&~  \, \fracm 12 \, (\g_a) {}_{\a}{}^{\b} \,
\r_{\b}(x) ~~~, \cr
[~ {\cal M}_a \, , \, A_b (x) \,  ] &=&~  \,
 \, \e_b{}_a{}_c
\,  A^c(x) ~~~.
\eea
Some useful identities:
\bea
X_{[\a\b ]}=-C_{\a\b}X^\g_{~\g}~~,~~\\
T_\g C_{\b\d}+T_\b C_{\d\g} + T_\d C_{\g\b}=0~~.
\eea


\subsection{Differential Forms }
\label{appa}

\setcounter{equation}{0}
\renewcommand{\theequation}{\thesubsection.\arabic{equation}}


$~~~$If $\alpha_p$ is a $p$-differential form then its expansion in
components is
\begin{equation}
\alpha_p={1 \over p!}\: \alpha_{m_1,\ldots,m_p} \: dx^{m_1}\wedge
\ldots \wedge dx^{m_p}\,.
\end{equation}
Let us consider the wedge product between a $p$-differential form
$\alpha_p$ and a $q$-differential form $\beta_q$.
$\alpha_p \wedge \beta_q$ is a $(p+q)$-differential form, so
\begin{eqnarray}
\alpha_p \wedge \beta_q = {1 \over \left(p+q\right)!} \: (\alpha_p
\wedge \beta_q)_{m_1,\ldots,m_{p+q}} \: dx^{m_1}\wedge \ldots
\wedge dx^{m_{p+q}}\,.
\end{eqnarray}
On the other hand, by definition
\begin{eqnarray}
\alpha_p \wedge \beta_q ={1 \over p!\, q!} \:
\alpha_{[m_1,\ldots,m_p} \, \beta_{m_{p+1},\ldots,m_{p+q}]} \:
dx^{m_1}\wedge \ldots \wedge dx^{m_{p+q}}\,,
\end{eqnarray}
therefore
\begin{eqnarray}
\label{3.7} \left( \alpha_p \wedge \beta_q
\right)_{m_1,\ldots,m_{p+q}} = {\left(p+q\right)! \over p! \, q!}
\: \alpha_{[m_1,\ldots,m_p} \, \beta_{m_{p+1},\ldots,m_{p+q}]}\,.
\end{eqnarray}
The definition for the exterior derivation is
\begin{equation}
d\alpha_p={1 \over p!} \: \partial_{[m_1} \,
\alpha_{m_2,\ldots,m_{p+1}]} \: dx^{m_1}\wedge \ldots \wedge
dx^{m_{p+1}}\,.
\end{equation}
But $d\alpha_p$ is a $(p+1)$-form
\begin{equation}
d\alpha_p={1 \over \left(p+1\right)!} \:
(d\alpha_p)_{m_1,\ldots,m_{p+1}} \: dx^{m_1}\wedge \ldots \wedge
dx^{m_p}\,,
\end{equation}
therefore
\begin{equation}
\label{3.13} \left( d\alpha_p \right)_{m_1,\ldots,m_{p+1}}= (p+1)
\:\partial_{[m_1}\, \alpha_{m_2,\ldots,m_{p+1}]}\,.
\end{equation}
The Hodge $\star$ operator of some $p\,$-form on a real
$n\,$-dimensional manifold is defined as
\begin{equation}
\star \, \alpha_p = {\sqrt{g} \over p!(n-p)!}\, \alpha_{k_1 \ldots
k_p} \, g^{k_1 m_1} \ldots g^{k_p m_p} \, \epsilon_{m_1 \ldots m_p
m_{p+1} \ldots m_n} \, dx^{m_{p+1}} \wedge \ldots \wedge
dx^{m_n}\,,
\end{equation}
where
\begin{equation}
\epsilon_{1 \ldots n} \,=\, +\,1\,.
\end{equation}
Regarding the integration of some $p\,$-form $\alpha_p$ on a
$p\,$-cycle ${\cal{C}}_p$ we have that
\begin{equation}
\int_{{\cal{C}}_p} \alpha_p = {1 \over p!} \int_{{\cal{C}}_p}
\alpha_{m_1 \ldots m_p} dx^{m_1} \wedge \ldots \wedge dx^{m_p}\,.
\end{equation}
We can also introduce an inner product on the space of real
$p\,$-forms defined on a $n\,$-dimensional manifold $\cal{M}$
\begin{equation}
\label{inner-product} \langle \alpha_p,\,\beta_p\rangle =
\int_{\cal{M}}\alpha_p \wedge \star \beta_p = {1 \over p!} \,
\int_{\cal{M}} \alpha_{m_1 \ldots m_p} \, \beta^{m_1 \ldots m_p}
\, \sqrt{g} \, dx^1 \wedge \ldots \wedge dx^n \,.
\end{equation}
%


\section{Derivation of Fierz Identities}
\label{fierz}

\setcounter{equation}{0}
\renewcommand{\theequation}{\thesection.\arabic{equation}}


$~~~$Choosing the real basis:
\bea
\g^0=i\s^2~~,~~\g^1=\s^1~~,~~\g^2=\s^3~~.
\eea
We can show by explicit substitution that:
\bea
\label{fierz1}
(\g^a)_{\a\b}(\g_a)^{\g\d}=-\d_{(\a}^{~~\g}\d_{\b)}^{~~\d}~~.
\eea
Basis free, we can derive:
\bea
(\g^a)_{\a\b}(\g_a)^{\g\d}&=&
(\g^a)_\a^{~\epsilon}(\g_a)_\eta^{~\d}C_{\epsilon\b}C^{\g\eta}=
(\g^a)_\a^{~\epsilon}(\g_a)_\eta^{~\d}\d_{[\epsilon}^{~\g}\d_{\b]}^{~\eta}\cr
&=&(\g^a)_\a^{~\g}(\g_a)_\b^{~\d}-\d_\b^{~\g}(\g^a\g_a)_\a^{~\d}\cr
&=&\frac 12 (\g^a)_{(\a}^{~~\g}(\g_a)_{\b)}^{~~\d}-\frac 32 \d_{(\a}^{~~\g}\d_{\b)}^{~~\d}~~.
\eea
Using this result and (\ref{fierz1}) we also have:
\bea
(\g^a)_{\a\b}(\g_a)^{\g\d}=-(\g^a)_{(\a}^{~~\g}(\g_a)_{\b)}^{~~\d}~~.
\eea
The second Fierz identity can be derived directly from the defining relation (\ref{gamma}):
\bea
\Big \{ (\g^{a})_{\g}^{~\s} (\g^{ b})_{\s}^{~\d} =
\eta^{ab} \d_{\g}^{~\d} +  \e^{abc}  (\g_{c})_{\g}^{~\d} \Big \}(\g_b)_{\a\b}~~.
\eea
Using (\ref{fierz1}) we can simplify this relation:
\bea
 \e^{abc}  (\g_b)_{\a\b}(\g_c)_{\g}^{~\d}&=&
(\g^a)_{\g\s}\d_{(\a}^{~~\s}\d_{\b)}^{~~\d}-(\g^a)_{\a\b}\d_\g^{~\g}\cr
&=&(\g^a)_{\a\g}\d_\b^{~\d}+(\g^a)_{\b[\g}\d_{\a]}^{~~\d}\cr
&=&(\g^a)_{\a\g}\d_\b^{~\d}+C_{\a\g}(\g^a)_\b^{~\d}~~.
\eea
A consequence of this identity is:
\bea
\label{cons}
(\g_{[c})_{(\a}^{~~\d}(\g_{d]})_{\b )}^{~~\s}=-2\e_{acd}C^{\d\s}(\g^a)_{\a\b}~~.
\eea


\section{Verification of Three-Dimensional Supergravity Covariant Derivative Algebra}
\label{3D-algebra}

\setcounter{equation}{0}
\renewcommand{\theequation}{\thesection.\arabic{equation}}


$~~~$The algebra of supergravity covariant derivatives given in
the literature is not written in our conventions, and does not
contain the gauge fields.  To get the correct algebra we take the
form given in the literature with arbitrary coefficients and add
the superfield strengths ${\cal F}_{\a\un b}$ and ${\cal F}^{\un c
I}$ associated with the $U(1)$ gauge theory: \bea [ \nabla_{\a}
~,~ \nabla_{\b} \}  &=&   (\g^{\un c})_{\a \b} ~ \nabla_{\un c}
~-~ (\g^{\un c})_{\a \b}R \, {\cal M}_{\un c} ~ ~~~,  \cr
~[ \nabla_\a ~,~ \nabla_{\un b} \}  &=&
~ -a(\g_{\un b})_{\a}{}^{\d} R
\nabla_{\d} ~+~[-2(\g_{\un b})_{\a}{}^{\d}  \S_{\d} {}^{\un d} ~+~ b \fracm 43 (\g_{\un b}\g^{\un d})_{\a}{}^{\e} ( \nabla_{\e} R )] {\cal M}_{\un d} ~\cr
&~&~+~c ( \nabla_{\a} R ) {\cal M}_{\un b}  + {\cal F}_{\a\un b}^It_I, \cr
[ \nabla_{\un a} ~,~ \nabla_{\un b} \}  &=&   +2
\e_{\un a \un b \un c } [~d \S^{\a \un c} + e \fracm 23 (\g^{\un
c})^{\a \b} (\nabla_{\b} R) ~] \nabla_{\a}  \cr
&~&+  \e_{\un a \un b \un c }[~  {\Hat {\cal R}}{}^{\un c \un d}
~+~ \fracm 23 \eta^{\un c \un d} (f\nabla^2 R ~+~ g\fracm 32 R^2 )~]
{\cal M}_{\un d}\cr
&~&+\e_{\un a\un b\un c}{\cal F}^{\un c I}t_I  ~~~,
\eea
where ${\Hat {\cal R}}{}^{a b} - {\Hat {\cal R}}{}^{b a} = \eta_{\un a
\un b} {\Hat {\cal R}}{}^{\un a \un b} = (\g_{\un d}
)^{\a \b} \S_{\b}{}^{\un d} = 0$ and
\bea
\label{curl}
\nabla_\a \S_\b^{~f} = -\frac 14 (\g^e)_{\a\b}\Hat{\cal R}_e^{~f}
+\frac 16[C_{\a\b}\h^{fd}+\frac 12 \e^{fde}(\g_e)_{\a\b}]\nabla_d R~~.
\eea
By checking the Bianchi identities,we will set the coefficients and derive constraints on the new superfield strengths as in (\ref{curl}).  The Bianchi identity $[[\nabla_{(\a} , \nabla_\b \} , \nabla_{\g)}\} =0$ looks like:
\bea
[[\nabla_{(\a} ,\nabla_\b \},\nabla_{\g)} \}=
-(\g^c)_{\a\b}\Big \{ (\frac 12 - a)(\g_c)_\g^{~\d}R\nabla_\d
+{\cal F}_{\g c}^It_I
+(c-1)(\nabla_\g R){\cal M}_c\cr
+(\g_c)_\g^{~\d}[-2\S_\d^d+\frac 43 b (\g^d)_\d^{~\e}(\nabla_\e R)]{\cal M}_d
\Big \}+[\b\g\a ]+[\g\a\b ]~~.
\eea
This equation is satisfied if $c=1$ and
\bea
(\g^{c})_{(\a\b}{\cal F}_{\g )c}^I=0~~~~
\Rightarrow~~~~{\cal F}_{\g c}^I={\frac 13}(\g_{c})_\g^{~\a}W_\a^I~~.
\eea
The identity $[\{\nabla_\a, \nabla_\b\} , \nabla_{c}\, ]
+\{ [\nabla_{c}, \nabla_{(\a}],\nabla_{\b)}\} =0$ is quite complicated, so we restrict our attention to one algebra element at a time.  The terms proportional to $t_I$ are:
\bea
(\g^d)_{\a\b}\e_{dce}{\cal F}^{e I}
-\frac 13 (\g_c)_{(\a}^{~~\d}\nabla_{\b )}W_{\d}^I=0~~.
\eea
Multiplying by $(\g^{c})^{\a\b}$ implies $\nabla^\d W_\d^I=0$. Multiplying by $(\g_a)^{\a\b}$ and antisymmetrizing over $a$ and $c$ leads to:
\bea
{\cal F}^{e I}={\frac 13}(\g^{e})_\b^{~\d}\nabla^\b W_\d^I~~.
\eea
Terms proportional to $\nabla_a$ are:
\bea
-(\g^d)_{\a\b}\e_{cde}R\nabla^e+a(\g_c)_{(\a}^{~~\d}(\g^d)_{\b )\d}R\nabla_d=0~~.
\eea
Which means $a=-\frac 12$.  Continuing to the terms proportional to $\S_{\b c}\nabla_\a$:
\bea
2d\e_{dce}(\g^d)_{\a\b}\S^{\d e}\nabla_\d
+(\g_c)_{(\a}^{~~\d}(\g_d)_{\b)}^{~~\s}\S_\d^d\nabla_\s=0~~.
\eea
Using (\ref{cons}) and the fact that $\S_{\a c}$ is gamma traceless, we see that $d=-1$.  The terms proportional to $R\nabla_\a$ are:
\bea
{\cal J}_{\a\b\g}\nabla^\g:=
\Big \{ \frac 43 e C_{\a\g}(\g_c)_b^{~\r}(\nabla_\r R)
+\frac 43 e (\g_c)_{\g\a}(\nabla_\b R
+(\g_c)_{\g(\a}(\nabla_{\b)}R)\cr
+\frac 43 b (\g_c)_{\a\b}(\nabla_\g R)
+\frac 23 b (\g_c)_{\g(\a}(\nabla_{\b)}R)
\Big \}~\nabla^\g~~.~~~
\eea
Therefore ${\cal J}_{\a\b\g}=0$.  ${\cal J}_{\a\b\g}$ is symmetric on $\a\b$ and therefore it is the sum of two independent irreducible spin tensors corresponding to the completely symmetric
$\tiny{\begin{array}{c} \fbox{}\fbox{}\fbox{}\\ \end{array} }$ and corner $\tiny{\begin{array}{c}\fbox{}\fbox{}\\ ~\,{\raise1.0ex\hbox{\fbox{}}} {~\,\,~~}\end{array} }$ tabluex.  Both of these should vanish separately.  Taking ${\cal J}_{(\a\b\g)}=0$ we see that:
\bea
4e+8b+6=0~~.
\eea
Then setting ${\cal J}^{\g}_{~\b\g}=0$ we have:
\bea
-8e+2b-3=0~~.
\eea
Thus,
\bea
e=b=-\frac 12~~.
\eea
We now turn to the last terms, they are proportional to the Lorentz generator.  Looking at non-linear terms involving $R$ we have:
\bea
\nonumber
[g-2a]\e_c^{~fd}(\g_d)_{\a\b}R^2{\cal M}_f =0 ~~,\cr
[f-4b]\frac 23 (\g^d)_{\a\b}\e_{dc}^{~~f}\nabla^2 R {\cal M}_f=0~~,\cr
\rightarrow ~~~~g=2a=-1~~~~~f=4b=-2~~.
\eea
Where we have used the
following fact to extract these contributions:
\bea
\nabla_\a\nabla_\b R = \frac 12 (\g^d)_{\a\b}(\nabla_d R) -
C_{\a\b}\nabla^2 R ~~.
\eea
The remaining terms in this Bianchi
identity are:
\bea \Big\{ (\g^d)_{\a\b}\e_{dce} \Hat{\cal R}^{ef}
+(\g^f)_{\a\b}(\nabla_c R)
+2(\g_c)_{(\a}^{~~\d}\nabla_{\b)}\S_\d^{~f}~~~~
~~~~~~~~~~~\cr
-\frac 23 b (\g_c\g^f\g^d)_{(\a\b)}(\nabla_d R) -(\g^d)_{\a\b}(\nabla_d
R)\d_c^{~f} \Big\}{\cal M}_f=0 ~~.
\eea
After converting the free
vector index into two symmetric spinor indices by contracting with
$(\g^c)_{\g\d}$ we have an expression of the form ${\cal
J}_{\a\b\g\d}^f{\cal M}_f=0$.  This tensor is the product of two
rank two symmetric spin tensors and has the following
decomposition in terms of tabluex:
$\tiny{\begin{array}{c}
\fbox{}\fbox{} \\ \end{array}}
\otimes
\tiny{\begin{array}{c} \fbox{}\fbox{} \\ \end{array}}
=\tiny{\begin{array}{c}\fbox{}\fbox{}\fbox{}\fbox{} \end{array}}
\oplus \tiny{\begin{array}{c}\fbox{}\fbox{}\\ {\raise1.0ex\hbox{\fbox{}\fbox{}}}
\end{array} }\oplus
\tiny{\begin{array}{c}\fbox{}\fbox{}\fbox{}\\
{\raise1.0ex\hbox{\fbox{}~~~~~\,}} \end{array}}$.  The completely symmetric term
vanishes identically.  The box diagram $\sim$
$C^{\g\a}C^{\d\b}{\cal J}_{\a\b\g\d}^f{\cal M}_f$ takes the form:
\bea
\nonumber 0=\{-2\nabla^c R -12\nabla^\d \S_\d^{~c}+8b\nabla^c
R+2\nabla^c R \}{\cal M}_c ~~.
\eea
Which implies that $\nabla^\s
\S_\s^{~f} = -\frac 13 \nabla^f R$.  The gun diagram $\sim$
$C^{\g\a}{\cal J}_{\a(\b\d)\g}^f{\cal M}_f$ takes the form:
\bea
\nonumber 0=\{-4(\g^e)_{\b\d}\Hat
{\cal R}_e^{~f}-8\nabla_{(\b}\S_{\d)}^{~~f} +(2+2-\frac
83)\e^{cde}(\g_e)_{\b\d}\nabla_d R \}{\cal M}_f ~~.
\eea
Which implies
that $\nabla_{(\b}\S_{\d)}^{~~f} = \frac 16
\e^{fde}(\g_e)_{\d\b}\nabla_d R -\frac 12 (\g^e)_{\b\d}\Hat
{\cal R}_e^{~f}$.  Thus, the spinorial derivative of $\S_\a^{~f}$ takes
the form:
\bea
\nabla_\a \S_\b^{~f} = -\frac 14 (\g^e)_{\a\b}\Hat
{\cal R}_e^{~f} +\frac 16[C_{\a\b}\h^{fd}+\frac 12
\e^{fde}(\g_e)_{\a\b}]\nabla_d R ~~.
\eea
This completes the analysis
of the spin-spin-vector Bianchi identity.  We now move on to the
spin-vector-vector Bianchi identity:
\bea
\nonumber [[\nabla_\a
,\nabla_b \} , \nabla_c\}+[[\nabla_b , \nabla_c \} , \nabla_\a \}
+[[\nabla_c , \nabla_\a \} , \nabla_b\}=0 ~~.
\eea
This identity is
satisfied identically, yielding no further constraints.  The final
identity is all vector derivatives: $[[\nabla_{[a} ,\nabla_b \} ,
\nabla_{c]} \}=0$.  This identity yields some more differential
constraints which are of no consequence to the derivations in the
body of this paper.


\section{Review of $Spin(7)$ Holonomy Manifolds}
\label{appendix-spin7}

\setcounter{equation}{0}
\renewcommand{\theequation}{\thesection.\arabic{equation}}


$~~~$This appendix contains a brief review of some of the relevant aspects of $Spin(7)$
holonomy manifolds. An elegant discussion can be found in \cite{Joyce}.
On an Riemannian manifold $X$ of dimension $n$, the spin connection ${\omega}$ is, in general,
an $SO(n)$ gauge field. If we parallel transport a spinor $\psi$ around a closed path $\gamma$,
the spinor comes back as $U\psi$, where $U=Pexp\int_{\gamma}{\omega}~dx$ is the path
ordered exponential of $\omega$ around the curve $\gamma$.

A compactification of ${\cal M}$-theory (or string theory) on $X$ preserves
some amount of supersymmetry if $X$ admits one (or more) covariantly constant spinors.
Such spinors return upon parallel transport to their original values, i.e. they satisfy $U\psi=\psi$.
The holonomy of the manifold is then a (proper) subgroup of $SO(n)$. A $Spin(7)$ holonomy manifold is an eight-dimensional manifold, for which one such spinor exists. Therefore, if we compactify
${\cal M}$-theory on these manifolds we obtain an ${\cal N}=1$ theory in three dimensions. $Spin(7)$ is a
subgroup of $GL(8,\IR)$ defined as follows. Introduce on ${\IR}^8$ the coordinates
$(x_1,\dots,x_8)$ and the four-form $dx_{ijkl}=dx_i\wedge dx_j\wedge dx_k\wedge dx_l$. Define a
self-dual 4-form  $\Omega$ on ${\IR}^8$ by
\begin{eqnarray}
\Omega &=& dx_{1234}+dx_{1256}+dx_{1278}+dx_{1357}-dx_{1368}  \nonumber \\
&-&dx_{1458}-dx_{1467}-dx_{2358}-dx_{2367}-dx_{2457} \\
&+&dx_{2468}+dx_{3456}+dx_{3478}+dx_{5678} \nonumber \,.
\end{eqnarray}
The subgroup of $GL(8,\IR)$ preserving $\Omega$ is the holonomy group
$Spin(7)$. It is a  compact, connected, simply connected, semisimple, twenty-one-dimensional Lie group, which is isomorphic to the  double cover of $SO(7)$.
Many of the mathematical properties of $Spin(7)$
holonomy  manifolds are discussed in detail in \cite{Joyce}. Let us here only mention that these manifolds
are Ricci flat  but are, in general, not K\"ahler.

The cohomology of a compact $Spin(7)$ holonomy
manifold can be  decomposed into the following representations of $Spin(7)$
\begin{eqnarray}
H^0(X, \IR) & = & \IR ~~,\nonumber \\
H^1(X, \IR) & = & 0 ~~,\nonumber \\
H^2(X, \IR) & = & H^2_{{\bf 21}}(X,\IR) ~~,\nonumber \\
H^3(X, \IR) & = & H^3_{{\bf 48}}(X, \IR) ~~,\nonumber \\
\label{cohomology}
H^4(X, \IR) & = & H^4_{{\bf 1}^+} (X, \IR) \oplus H^4_{{\bf 27}^+}
(X, \IR) \oplus H^4_{{\bf 35}^-} (X, \IR) ~~,\\
H^5 (X, \IR) & = & H^5_{{\bf 48}} (X, \IR) ~~,\nonumber \\
H^6 (X, \IR) & = & H^6_{{\bf 21}} (X, \IR) ~~,\nonumber \\
H^7 (X, \IR) & = & 0 \nonumber ~~,\\
H^8 (X,\IR) & = &\IR \nonumber~~.
\end{eqnarray}
where the label $``\pm"$ indicates self-dual and anti-self-dual
four-forms respectively and the  subindex indicates the
representation. The Cayley calibration $\Omega$ belongs to the
cohomology $H_{{\bf 1}^+}^4(X,\IR)$.

Next we will briefly discuss deformations of the Cayley
calibration. More details can be found in \cite{Joyce} and
\cite{Karigiannis}. The tangent space to the family of torsion-free
$Spin(7)$ structures, up to diffeomorphism is naturally isomorphic
to the direct sum $H^4_{{{\bf 1}^+}}(X, \IR) \oplus H^4_{{\bf
35}^-} (X, \IR)$ if $X$ is compact and the holonomy is $Spin(7)$ and not some proper
subgroup. Thus, if the holonomy is $Spin(7)$ the family has
dimension $1+b_4^-$, and the infinitesimal variations in $\Omega$
are of the form $c\Omega+ \xi$ where $\xi$ a harmonic
anti-self-dual four-form and $c$ is a number.

When we are moving in moduli space
along the ``radial direction'' $\phi$, the Cayley calibration
deformation is
\begin{equation}
\delta\Omega=K \delta \phi \Omega\,,
\end{equation}
or in other words
\begin{equation}
\label{partial-0-omega} {\partial \Omega \over \partial \phi}= K
\Omega \,.
\end{equation}
If we consider infinitesimal displacements in moduli space along
the other $b_4^-$ directions, then the Cayley calibration
deformation is
\begin{equation}
\delta\Omega=\delta \phi^A(\xi_A-K_A \Omega) \,,
\end{equation}
or in other words
\begin{equation}
\label{partail-a-omega} {\partial \Omega \over \partial
\phi^A}=\xi_A-K_A \Omega \,,
\end{equation}
where $\xi_A$ are the anti-self-dual harmonic four-forms.  If the
movement in the moduli space is not along some particular
direction then
\begin{equation}
\delta\Omega=\delta \phi^A \, \xi_A +(\delta \phi \, K - \delta \phi^A \, K_A )\Omega \,.
\end{equation}
We note that the potential
\begin{equation}
P={1 \over 2}\ln(\int_{M_8} \Omega \wedge \star \Omega) \,,
\end{equation}
generates $K=\partial P$ and $K_A=-\partial_A P$. The fact that
\begin{equation}
\int_{M_8} \Omega \wedge \star \Omega = 14 {\cal V}_{M_8} = e^{2P} \,,
\end{equation}
fixes $K=2$, where ${\cal V}_{M_8}$ is the volume of the internal manifold.


\section{Dimensional Reduction of the Einstein-Hilbert Term}
\label{appendix-dimensional}

\setcounter{equation}{0}
\renewcommand{\theequation}{\thesection.\arabic{equation}}


$~~~$In this appendix we present the technical details related to the
compactification of the Einstein-Hilbert term. We treat first the
zero flux case and then we calculate the reduction for the
non-zero background flux case. As usual the Greek indices refer
to the external space, the small Latin indices refer to the
internal space,  and finally the capital Latin indices refer to
the entire eleven dimensional space.


\subsection{Zero Flux Case}
\label{appendix-zero-flux}

\setcounter{equation}{0}
\renewcommand{\theequation}{\thesubsection.\arabic{equation}}


$~~~$We start with the following ansatz for the inverse metric
\begin{equation}
g^{mn}(x,y)= \hat{g}^{mn}(y) + \phi(x) \hat{g}^{mn}(y) + \sum_{A=1}^{b_4^-}\phi^A(x) ~ h^{mn}_A(x,y) + \ldots \,,
\end{equation}
where we have denoted by $g^{mn}(x,y)$ the inverse metric of
$g_{mn}(x,y)$ and by $\hat{g}^{mn}(x)$ the inverse metric of
$\hat{g}_{mn}(x)$
\begin{equation}
g_{mn}(x,y) \, g^{np}(x,y)={\delta_m}^p ~~,~~
\hat{g}_{mn}(y) \, \hat{g}^{np}(y)={\delta_m}^p\,.
\end{equation}
Due to these facts we obtain that
\begin{equation}
h^{mn}_A(x,y)=-\hat{g}^{ma}(y) \, e_{A\,ab}(y) \, \hat{g}^{bn}(y)\,.
\end{equation}
The tracelessness of $e_{A\,ab}$ implies the tracelessness of
$h^{mn}_A$.  The ansatz (\ref{g-ansatz}) implies that the only non-zero Christoffel symbols are
\begin{eqnarray}
\label{Christoffel} &&\Gamma^\alpha_{\mu \nu} = {1 \over
2}g^{\alpha \beta} \,(\partial_\mu g_{\beta \nu} + \partial_\nu
g_{\mu \beta} - \partial_\beta
g_{\mu \nu})\,,\nonumber\\&&\Gamma^\alpha_{m \nu} = 0\,,\nonumber\\
&&\Gamma^\alpha_{mn} = -\,{1 \over 2}g^{\alpha \beta}
\,(\partial_\beta g_{mn})\,,\nonumber\\
&&\Gamma^a_{mn} = {1 \over 2}g^{ab} \, (\partial_m g_{bn} +
\partial_n g_{mb} - \partial_b g_{mn})\,,\nonumber\\
&&\Gamma^a_{\mu n} = {1 \over 2}g^{ab} \, (\partial_\mu g_{bn})\,,\nonumber\\
&&\Gamma^a_{\mu \nu} = 0\,.
\end{eqnarray}
Using the following definition of the Ricci tensor
\begin{equation}
R^{(11)}_{MN}=\partial_A \Gamma^A_{MN} - \partial_N \Gamma^A_{MA}
+ \Gamma^A_{MN}\Gamma^B_{AB} - \Gamma^A_{MB}\Gamma^B_{AN}\,,
\end{equation}
we can derive the expression for the eleven-dimensional Ricci
scalar
\begin{eqnarray}
\label{Ricci} R^{(11)} &=& R^{(3)} + R^{(8)} + g^{mn}
\partial_\alpha \Gamma_{mn}^\alpha - g^{\mu \nu} \partial_\nu
\Gamma^a_{\mu a} + g^{\mu \nu} \Gamma^\alpha_{\mu \nu}
\Gamma^b_{\alpha b}
 + g^{mn} \Gamma^\alpha_{mn} \Gamma^\beta_{\alpha \beta} \nonumber \\
&+&\, g^{mn} \Gamma^\alpha_{mn} \Gamma^b_{\alpha b} - \left[
g^{\mu \nu} \Gamma_{\mu b}^a \Gamma_{a \nu}^b + g^{mn} \Gamma_{m
\beta}^a \Gamma_{an}^\beta + g^{mn} \Gamma_{mb}^\alpha
\Gamma_{\alpha n}^b \right] \,.\end{eqnarray}
where $R^{(3)}$ is the three-dimensional Ricci scalar and
$R^{(8)}$ is the eight-dimensional Ricci scalar. We can determine
that
\begin{eqnarray}
\label{integrate-Ricci} \int_{\cal M} d^{11}x \, \sqrt{-g^{(11)}}
\, R^{(11)} &=& \int_{\cal M} d^{11}x \, \sqrt{-g^{(11)}} \,
\Big\{ R^{(3)} + g^{\mu \nu} \Gamma_{\mu a}^a \Gamma_{\nu b}^b -
(\partial_\alpha g^{mn}) \Gamma_{mn}^\alpha   \nonumber\\&-&\,
\left[  g^{\mu \nu} \Gamma_{\mu b}^a \Gamma_{\nu a}^b +g^{mn}
\Gamma_{\beta m}^a \Gamma_{an}^\beta + g^{mn} \Gamma_{mb}^\alpha
\Gamma_{\alpha n}^b \right] \Big\} \,,
\end{eqnarray}
where we have integrated by parts with respect to the internal
coordinates and we have used the fact that the internal manifold
is Ricci flat, i.e. $R^{(8)}=0\,$.  It is easy to see that
\begin{eqnarray}
&&\int_{\cal M} d^{11}x \sqrt{-g^{(11)}} \, g^{\mu \nu} \,
\Gamma_{\mu a}^a \, \Gamma_{\nu b}^b = {\cal V}_{M_8} \int_{M_3}
d^3x \, \sqrt{-g^{(3)}} \, \Big\{\,16\,(\partial_\alpha \phi) \,
(\partial^\alpha \phi)\Big\} \,,
\end{eqnarray}
\begin{eqnarray}
&&\int_{\cal M} d^{11}x \sqrt{-g^{(11)}} \, (\partial_\alpha
g^{mn}) \Gamma_{mn}^\alpha = \,2\, {\cal V}_{M_8} \int_{M_3} d^3 x
\,\sqrt{-g^{(3)}}\, \Big\{ \,2\,(\partial_\alpha \phi) \, (\partial^\alpha \phi) \nonumber \\
&&+\, \sum_{A,B=1}^{b_4^-} \, {\cal L}_{AB} \, (\partial_\alpha
\phi^A) \, (\partial^\alpha \phi^B) \Big\}\,,
\end{eqnarray}
\begin{eqnarray}
&&\int_{\cal M} d^{11}x \sqrt{-g^{(11)}} \, \left[  g^{\mu \nu}
\Gamma_{\mu b}^a \Gamma_{\nu a}^b + g^{mn} \Gamma_{\beta m}^a
\Gamma_{an}^\beta +
g^{mn} \Gamma_{mb}^\alpha \Gamma_{\alpha n}^b \right] \nonumber \\
&& =\,-\,{\cal V}_{M_8}\, \int_{M_3} d^3x \sqrt{-g^{(3)}}\,\Big\{
\,2\,(\partial_\alpha \phi) \, (\partial^\alpha \phi) \nonumber \\
&&+ \,\sum_{A,B=1}^{b_4^-} \, {\cal L}_{AB} (\partial_\alpha \phi^A) \,
(\partial^\alpha \phi^B) \Big\}\,,
\end{eqnarray}
where ${\cal L}_{AB}$ was defined in (\ref{moduli-metric-l}) and
${\cal V}_{M_8}$ represents the volume of the internal manifold and it
is defined in (\ref{volume}).

We know that after compactification we arrive in the string frame
even if we started in eleven dimensions in the Einstein frame.
Therefore we have to perform a Weyl transformation for the
external metric. The fact that we do not see any exponential of
the radial modulus seating in front of $R^{(3)}$ is because we
have consistently neglected higher order contributions in moduli
fields. However it is not difficult to realize that the Weyl
transformation that has to be performed is
\begin{equation}
g_{\alpha \beta} \rightarrow e^{-8 \phi}~g_{\alpha \beta} \,.
\end{equation}
The only visible change in this order of approximation is the
coefficient in front of the kinetic term for radion. All the other
terms in the action remain unchanged. Therefore the
Einstein-Hilbert term is
\begin{eqnarray}
\label{Einstein-Hilbert}
{1 \over 2 \kappa_{11}^2} \int_{\cal M} d^{11}x
\,\sqrt{-g^{(11)}}\, R^{(11)} &=& {1 \over 2\kappa_3^2} \, \int_{M_3}
d^3x \,\sqrt{-g^{(3)}}\,  \Big\{ R^{(3)} -
18(\partial_\alpha \phi) \, (\partial^\alpha \phi)  \nonumber\\
&&-\, \sum_{A,B=1}^{b_4^-} \, {\cal L}_{AB} \,(\partial_\alpha
\phi^A) \, (\partial^\alpha \phi^B)\Big\} + \ldots \,.
\end{eqnarray}
%


\subsection{Non-Zero Flux Case}
\label{appendix-non-zero-flux}

\setcounter{equation}{0}
\renewcommand{\theequation}{\thesubsection.\arabic{equation}}


$~~~$It is easy to derive an expression for the Ricci scalar in the
non-zero background case. For this task we rewrite the warped
metric (\ref{warped-metric}) as
\begin{equation}
\label{conformal-transformation} \widetilde{g}_{MN}=\Omega^2(y)
\, \bar{g}_{MN} \,,
\end{equation}
where $\Omega(y)=e^{\Delta(y)/3}$ and therefore
\begin{equation}
\label{conformal-metric} \bar{g}^{(11)}_{MN}(x,y) = \left(
\begin{array}{cc}
                     {g}^{(3)}_{\mu \nu}(x) & 0 \\
                     0 & \Omega^{-3}(y) \, {g}^{(8)}_{mn}(x,y)
                     \end{array} \right) \,.
\end{equation}
The Christoffel symbols that correspond to the metric
(\ref{conformal-metric}) are
\begin{eqnarray}
\label{conformal-Christoffel}
&&\bar{\Gamma}^\alpha_{\mu \nu} = \Gamma^\alpha_{\mu \nu} \,, \nonumber\\
&&\bar{\Gamma}^\alpha_{m \nu} = \Gamma^\alpha_{m \nu} \,,\nonumber\\
&&\bar{\Gamma}^\alpha_{mn} = \Omega^{-3} \, \Gamma^\alpha_{mn} \,,\nonumber\\
&&\bar{\Gamma}^a_{mn} = \Gamma^a_{mn} - {3 \over 2} \, \left[
\delta^a_m \partial_n +
\delta^a_n \partial_m - g_{mn} g^{ab}\partial_b \right] \, \ln \Omega \,,\nonumber\\
&&\bar{\Gamma}^a_{\mu n} = \Gamma^a_{\mu n} \,,\nonumber\\
&&\bar{\Gamma}^a_{\mu \nu} = \Gamma^a_{\mu \nu} \,,
\end{eqnarray}
where the unbarred symbols are computed in (\ref{Christoffel}). We
can repeat the computation for the Ricci scalar that correspond to
the metric (\ref{conformal-metric}) and at the end we will obtain
a similar formula to (\ref{Ricci}).  Due to the simple relations
(\ref{conformal-Christoffel}) between the Christoffel symbols, the
Ricci scalar for the metric (\ref{conformal-metric}) reduces to
\begin{equation}
\label{conformal-Ricci} \bar{R}^{(11)} = R^{(11)} + 21 \,
\Omega^3 \left[ g^{ab}\nabla_a \nabla_b \ln \Omega - {9 \over 2}
\, g^{ab} \, \nabla_a \ln \Omega \, \nabla_b \ln \Omega \right]
\,,
\end{equation}
where ${R}^{(11)}$ is given in (\ref{Ricci}).

To compute the Ricci scalar that corresponds to the metric
(\ref{warped-metric}) we have to perform the conformal
transformation (\ref{conformal-transformation}). The result
expressed in terms of the warp factor $\Delta(y)$ is
\begin{equation}
\label{warped-Ricci} \widetilde{R}^{(11)} = e^{-2\Delta(y)/3}
R^{(11)} + e^{\Delta(y)/3} \, \left[ \, {1 \over 3} \,
g^{ab}\nabla_a \nabla_b \Delta(y) - {1 \over 2} \, g^{ab} \,
\nabla_a \Delta(y) \, \nabla_b \Delta(y) \, \right] \,.
\end{equation}
Using the fact that the second term in (\ref{warped-Ricci})
produces a total derivative term which vanishes by Stokes'
theorem and the last term is subleading, we obtain that
\begin{equation}
\int_{\cal M} \, d^{11}x \,\sqrt{-\widetilde{g}^{(11)}}\,
\widetilde{R}^{(11)} \, = \, \int_{\cal M} d^{11}x
\,\sqrt{-g}\, R^{(11)} \, e^{- \Delta(y)} + \ldots \,.
\end{equation}
As expected, to leading order the kinetic coefficients receive no corrections from warping. Therefore we conclude that
\begin{eqnarray}
\label{Einstein-Hilbert-warped}
{1 \over 2 \kappa_{11}^2} \int_{\cal M} d^{11}x
\,\sqrt{-\widetilde{g}^{(11)}}\, \widetilde{R}^{(11)} &=& {1 \over 2\kappa_3^2} \, \int_{M_3}
d^3x \,\sqrt{-g^{(3)}}\,  \Big\{ R^{(3)} -
18(\partial_\alpha \phi) \, (\partial^\alpha \phi)  \nonumber\\
&&-\, \sum_{A,B=1}^{b_4^-} \, {\cal L}_{AB} \,(\partial_\alpha
\phi^A) \, (\partial^\alpha \phi^B)\Big\} + \ldots \,.
\end{eqnarray}
%


\pagebreak


\bibliographystyle{hunsrt}
\bibliography{bibliography}


\end{document}